# Elasto-visco-plastic flows in benchmark geometries:
# I. 4 to 1 Planar Contraction

*Milad Mousavi, Yannis Dimakopoulos, John Tsamopoulos\**

Laboratory of Fluid Mechanics and Rheology, Dep. of Chemical Engineering

University of Patras, Greece, 26504

## *Abstract*

We present predictions for the flow of elastoviscoplastic (EVP) fluids in the 4 to 1 planar contraction geometry. The Saramito-Herschel-Bulkley fluid model is solved via the finite-volume method with the OpenFOAM software. Both the constitutive model and the solution method require using transient simulations. In this benchmark geometry, whereas viscoelastic fluids may exhibit two vortices, referred to as lip and corner vortices, we find that EVP materials are unyielded in the concave corners. They are also unyielded along the mid-plane of both channels, but not around the contraction area where all stress components are larger. The unyielded areas using this EVP model are qualitatively similar to those using the standard viscoplastic models, when the Bingham or the Weissenberg numbers are lower than critical values, and then, a steady state is reached. When these two dimensionless numbers increase while they remain below the respective critical values, which are interdependent, (a) the unyielded regions expand and shift in the flow direction, and (b) the maximum velocity increases at the entrance of the contraction. Increasing material elasticity collaborates with increasing the yield stress, which expands the unyielded areas, because it deforms the material more prior to yielding compared to stiffer materials. Above the critical Weissenberg number, transient variations appear for longer times in all variables, including the yield surface, instead of a monotonic approach to the steady state. They may lead to oscillations which are damped or of constant amplitude or approach a flow with rather smooth path lines but complex stress field without a plane of symmetry, under creeping conditions. These patterns arise near the entrance of the narrow channel, where the curvature of the path lines is highest and its coupling with the increased elasticity triggers a purely elastic instability. Similarly, a critical value of the yield stress exists above which such phenomena are predicted.

# 1  Introduction

Understanding the behavior of non-Newtonian fluids in prototype geometries can enhance the efficiency and productivity of industrial equipment. Flow through a channel contraction or an expansion, or a contraction followed by an expansion, are prototype flows for many popular industrial applications, such as the production of plastic pipes, films, and other more complicated objects of different geometries; the flow in oil reservoirs or other porous media; the filling of bottles and cans with foodstuffs or personal care creams and medicine, and so on [1].

To identify the origins of difficulties in numerical simulations, researchers have proposed various sets of simple geometries as reference flows [2, 3]. The contraction geometry has been used to study different flow structures of viscoplastic and viscoelastic materials with different constitutive models. The same effort has not been undertaken with elasto-visco-plastic materials, which combine the properties of the above two major categories of non-Newtonian fluids. Such materials are foams, creams, gels, emulsions, foodstuffs, crude oil, blood (accounting for plasticity and elasticity generated by cell aggregation and fibrinogen), etc. This is one of the goals of this paper.

Out of the numerous experimental studies on the contraction flow of viscoelastic fluids, one should mention the two by Rodd et al. [4, 5], who investigated the flow of dilute polyethylene oxide (PEO) over a wide range of the relevant dimensionless numbers, Weissenberg ($Wi$) and Reynolds ($Re$). They presented a phase diagram on a $Wi - Re$ plane with different flow regimes, such as Newtonian-like, steady, unsteady, vortex growth regimes, etc. In the flow of a viscoelastic fluid through the sudden contraction both shear and extensional components of the stress arise [2]; the former dominate near the walls, while the latter are intensified at the center plane near the contraction [6]. A singular point appears at the re-entrant corner [6, 7]. The occurrence of vortices close to the re-entrant and concave corners is one of the most significant flow characteristics observed in both experiments and numerical simulations. These flow structures are the so-called lip and corner vortices, respectively. The appearance of these vortices strongly depends on the rheological properties of the fluid as well as geometrical characteristics [8].

Alves et al. [7] conducted a comparative analysis in a 4 to 1 planar contraction flow using different viscoelastic models. The shear-thinning in both the shear viscosity and the first normal stress difference of a PTT fluid allowed reaching much higher Weissenberg numbers with this model rather than with the Oldroyd-B model. Additionally, the intensity and size of the corner vortex was greater in the PTT model compared to the Oldroyd-B model. The study also highlighted that numerical convergence could occur with coarser meshes using the PTT model. The most significant observation was the presence of the lip vortex close to the re-entrant corners exclusively in the Oldroyd-B model, in which the extensional viscosity becomes unbounded above a certain extension rate.

Even 3D numerical investigations [9] have been conducted using the finite volume method (FVM) and the Oldroyd-B model. The results of this study revealed that as $Wi$ increased, the size of the corner vortex decreased. Nevertheless, the prediction of the first normal stress difference in the vicinity of the re-entrant corners was lower than the experimental data due to less accurate experimental characterization of the fluid in extensional flow. Additionally, the lip vortex was not observed across a wide range of Weissenberg and Reynolds numbers.

In another investigation, both numerical simulations using the PTT viscoelastic model and experimental visualizations were conducted in contraction flow [10]. The numerical simulations confirmed the existence of lip vortices near the re-entrant corners at extremely low Weissenberg numbers ($Wi < 1$). However, capturing the presence of the lip vortex experimentally posed challenges due to difficulties in visualization at low flow rates. Furthermore, the study examined the effect of the contraction ratio ($CR$), defined as the ratio of the wide channel width to the narrow channel width. It was observed that for all $CRs$, the size of the corner vortex increased as the Weissenberg number increased. Additionally, the velocity overshoot at the contraction entrance was absent in simulations with a shear thinning model, indicating that the presence of elasticity is necessary to observe this particular behavior.

The shape and extent of yielded/unyielded regions for a 4:1 planar contraction of a Bingham plastic have been reviewed in [11]. They noted two plug regions in the upstream and downstream channels, which are truly unyielded regions (TUR) and move with a plug velocity profile and two apparently unyielded regions (AUR) in the salient corners, where the velocities are very small, and the fluid in the area behaves as nearly stagnant and non-deforming. Due to the Papanastasiou regularization, which turns the "solid" region into a high-viscosity fluid [12], there are very small velocities and velocity gradients there, and the stresses remain below the yield stress. Similar 4:1 contraction flow takes place in the gas-assisted displacement of viscoplastic fluids from a 4:1 planar contraction and from a complex tube [13, 14].

Furthermore, in [15, 16], viscoplastic materials were studied through a 1 to 4 expansion flow with the Herschel-Bulkley & Bingham models. According to their findings, yield stress and inertia forces affect the flow in opposite directions. The reduction of the Herschel-Bulkley index "n" decreases the dimensions of the unyielded zones [16]. Alexandrou et al. [17] used the Herschel-Bulkley viscoplastic model with the Papanastasiou regularization in three-dimensional expansion flows. They showed that the structure of the material strongly depends on the Reynolds and Bingham numbers and the exponent $n$. This dependence is more visible at the concave corners, where the unyielded regions appear.

Until very recently, elasticity was excluded when yield-stress materials were investigated. However, this led to the inability to explain and predict many experimental observations with yield-stress materials, such as the loss of fore-and-aft symmetry under creeping flow conditions and the occurrence of the negative wake structure behind a sedimenting spherical particle through a Carbopol gel. In the case of a bubble rising through a yield-stress material, the negative wake structure was reported, and even more intriguingly, the bubble attained an inverted teardrop shape [18, 19]. These experimental observations were clear manifestations of elasticity in yield stress materials. To overcome the deficiency of viscoplastic models, Fraggedakis et al. [20] used the elastoviscoplastic (EVP) model proposed by Saramito [21], which combined the Bingham with the Oldroyd-B fluid models. They predicted accurately the loss of the fore-and-aft symmetry and the negative wake behind a solid particle. Subsequently, Saramito proposed a new model combining the Herschel-Bulkley with the Oldroyd fluid models [22]. Among others, this model has been used by Varchanis et al. [23] who got excellent agreement between experiments with a Pluronic aqueous solution in the optimized shape cross-slot extensional rheometer (OSCER) and simulations and Moschopoulos et al. [24] who captured the inverted teardrop shape of a bubble rising in Carbopol solutions.

This paper reports only the first part of our recent presentation at the International Rheology Conference (ICR) [25]. It examines the flow in contraction geometry of a fluid following the Saramito-Herschel-Bulkley (SHB) model. First, the numerical method (OpenFOAM) and our formulation are validated by reproducing results from the literature in the same geometry using the Oldroyd-B fluid model, which is the most challenging among other viscoelastic models. Both the SHB model and the solution software require transient simulations, even when a steady state is reached. In the contraction flow, the size of the two existing vortices is found to be highly affected by both the mesh quality and $Wi$. Consequently, the size of these vortices serves as a suitable parameter for validating and comparing the numerical results. Subsequently, the rheological parameters of the Carbopol-based EVP materials are determined through fitting experimental data [18, 19]. These materials are three quite similar Carbopol solutions, which are used as prototypes for simulations in this geometry. Subsequently, the study explores the effect of the rheological parameters of the SHB model and the relevant non-dimensional groups, on the yielded/unyielded areas. The elastic modulus and the yield stress are the parameters causing the most drastic changes and are examined in more detail. Especially when they exceed certain values, the flow may remain transient for a long time or even elastic turbulence may appear although the Reynolds number is much smaller than one.

## 2 Problem formulation

The governing equations in dimensional form under the assumptions of isothermal and incompressible flow and in the absence of body forces are the mass (eq. (1)) and momentum balances (eq. (2)):

$$\nabla \cdot \boldsymbol{u} = 0 \tag{1}$$

$$\rho \left(\frac{\partial \boldsymbol{u}}{\partial t} + \boldsymbol{u} \cdot \nabla \boldsymbol{u}\right) = -\nabla P + \nabla \cdot \boldsymbol{\tau}' \tag{2}$$

where the extra stress, $\boldsymbol{\tau}'$, is divided in the contributions from the solvent, $\boldsymbol{\tau}_s$, and the yield material, $\boldsymbol{\tau}$, and a bold font indicates a vectorial or tensorial variable. The constitutive equation of the latter is given by the thermodynamically consistent Saramito-Herschel-Bulkley model [22]:

$$\boldsymbol{\tau}' = \boldsymbol{\tau}_s + \boldsymbol{\tau} \tag{3}$$

$$\boldsymbol{\tau}_s = \eta_s (\nabla \boldsymbol{u} + (\nabla \boldsymbol{u})^T) \tag{4}$$

$$\frac{1}{G}\overset{\nabla}{\boldsymbol{\tau}} + \max\left(0, \frac{\|\boldsymbol{\tau}_d\| - \tau_0}{k\|\boldsymbol{\tau}_d\|^n}\right)^{1/n} \boldsymbol{\tau} = (\nabla \boldsymbol{u} + (\nabla \boldsymbol{u})^T) \tag{5}$$

In eq. (5), the first term in the left-hand-side is the elastic contribution and the second one the viscoplastic contribution to the rate-of-strain tensor, the max-term imposes the von Mises criterion for material yielding and the term $\overset{\nabla}{\boldsymbol{\tau}}$ indicates the upper-convected time derivative.

$$\overset{\nabla}{\boldsymbol{\tau}} = \frac{\partial \boldsymbol{\tau}}{\partial t} + \boldsymbol{u} \cdot \nabla \boldsymbol{\tau} - \boldsymbol{\tau} \cdot \nabla \boldsymbol{u} - (\nabla \boldsymbol{u})^T \cdot \boldsymbol{\tau} \tag{6}$$

Also, the deviatoric stress tensor, its second invariant and its trace, respectively, are defined in cartesian geometry as:

$$\boldsymbol{\tau}_d = \boldsymbol{\tau} - \frac{\text{tr}(\boldsymbol{\tau})}{N}\boldsymbol{I}, \; \|\boldsymbol{\tau}_d\| = \sqrt{0.5(\boldsymbol{\tau}_d : \boldsymbol{\tau}_d)}, \; \text{tr}(\boldsymbol{\tau}) = \tau_{xx} + \tau_{yy} + \tau_{zz} \tag{7}$$

In the equations above, $N$ is the number of the dimensions of the problem geometry, $\eta_s$ the solvent viscosity, $\eta_p$ the zero-shear polymeric viscosity, $\rho$ the fluid density, $G$ the elastic modulus, $k$ the consistency constant, $\tau_0$ the yield stress, $n$ the Herschel-Bulkley index, $(\nabla \boldsymbol{u})^T$ the transpose of the velocity gradient tensor, and $\boldsymbol{I}$ the unit tensor. Below yielding, the max term in eq. (5) is zero, and the hyperelastic model for small solid deformations is recovered; when $G$ approaches infinity, the Herschel-Bulkley model is recovered.

Although the simulations are carried out in dimensional form, it is easier to understand and explain the predictions by referring to the relevant five dimensionless groups:

$$Re = \frac{\rho U w}{\eta_0}, \; Wi = \frac{\lambda U}{w}, \; Bn = \frac{\tau_0 w}{\eta_0 U}, \; \beta = \frac{\eta_s}{\eta_0}, \; \varepsilon_y = \frac{\tau_0}{G}, \tag{8}$$

Where $U$ is the average velocity at the exit of the narrow channel, see Fig. 1, and the relaxation time and the total material viscosity are defined, respectively, as:

$$\lambda = \eta_p/G, \text{ and } \eta_0 = \eta_s + \eta_p = \eta_s + k(\frac{w}{U})^{1-n} \tag{9}$$

The first dimensionless number is the Reynolds number, $Re$, which is the ratio of inertial to viscous forces; the Weissenberg number, $Wi$, is the ratio of relaxation time to a characteristic shear rate; the Bingham number, $Bn$, is the ratio of yield stress to the viscous stress; $\beta$ is the ratio of the solvent viscosity to the total viscosity, and the yield strain, $\varepsilon_y$, is the ratio of the yield stress to the elastic modulus. When the solvent viscosity is negligible, as it is here, the yield strain is the product of the Weissenberg and the Bingham numbers, $\varepsilon_y = Wi * Bn$.

## 2.1 Geometry and mesh details

The top panel in Fig. 1 depicts the geometry of the abrupt 4 to 1 planar contraction flow, including typical lengths, and indicatively the size and the location of possible vortices in the inset. The three lower panels in Fig. 1 give certain details of the meshes VE M3 and EVP M4 (to be defined in table 1 and 4, respectively) close to the contraction and the re-entrant corner.

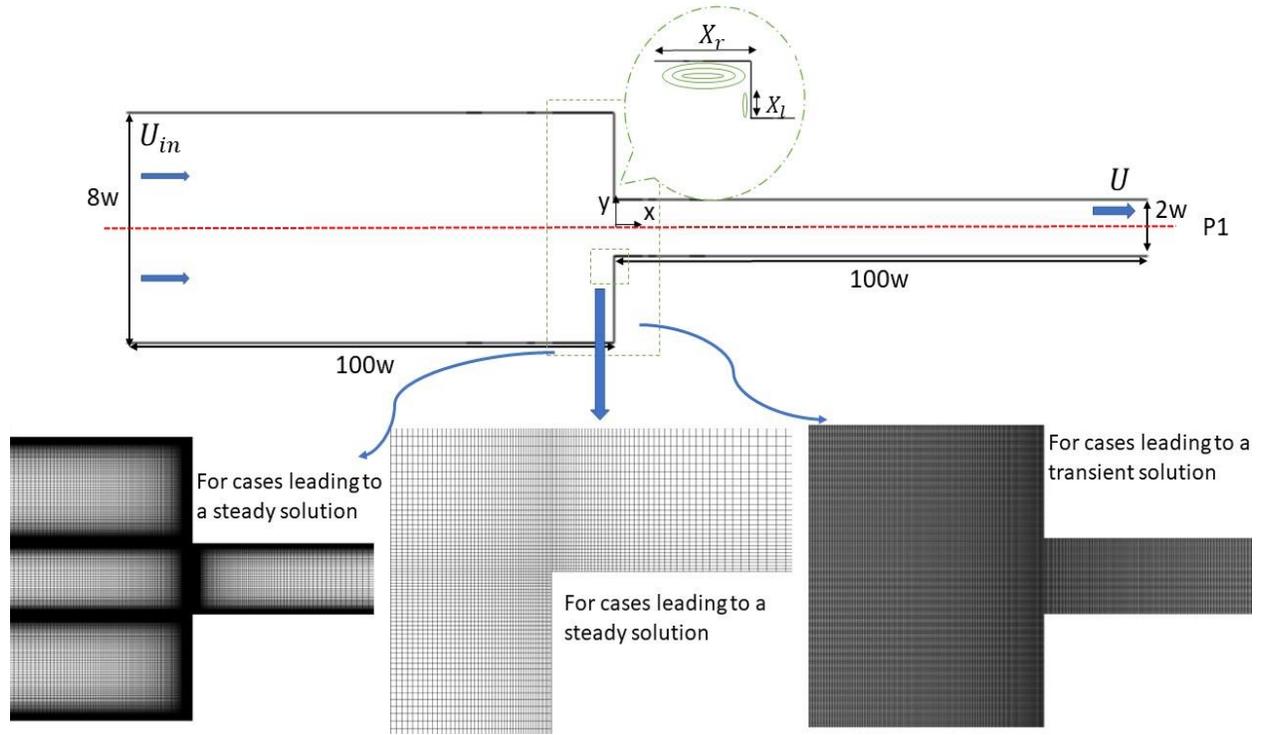

Figure 1. Flow geometry of the 4:1 contraction flow. The inset of the top panel indicates the location of the possible lip and corner vortices. The lower-left panel depicts a blow up of a mesh used for viscoelastic simulations (VE M3) around the contraction area and the lower-mid panel depicts a blow up of this mesh (VE M3) around the re-entrant corner. For cases leading to the transient solution with EVP fluids the lower-right panel depicts a blow up of the mesh (EVP M4).

The characteristic length for this geometry is half the width, $w$, of the narrow channel. The width of the wide channel is $8w$. The length shown for either channel is $100w$. We have determined that with this length fully developed flow is achieved well before the contraction and well before the outflow boundary and that inflow and outflow boundary conditions do not interfere with the flow around the contraction.

At the inlet, we apply a uniform velocity profile $U_{in}$, zero extra-stress components, and zero-gradient for the pressure. At the outlet, we consider the flow to be fully developed. Therefore, we apply a zero-gradient for every variable, except for the pressure which is fixed at zero (*p*=0). On the channel walls, we apply the no-slip, no-penetration boundary conditions for the velocity and a zero-gradient for the pressure. Also, we use a linear extrapolation of the stress components to increase the accuracy of the numerical simulation as suggested in [6].

In all viscoelastic simulations (performed to validate the numerical method of solution), $\beta$ is 1/9. The Weissenberg number varies from 0 (Newtonian fluid) to 12. The flow is assumed to be creeping ($Re < 1$). A mesh independence study was conducted first to determine its size necessary to achieve numerical accuracy at a reasonable computational cost. High spatial resolution is sought especially near the re-entrant corners to capture any complications that may arise in the fluid flow more accurately (lip and salient corner vortices). Mesh refinement is undertaken around the singular points (e.g. re-entrant and salient corners and contraction plane), but the mesh resolution decreases as the inlet or outlet boundaries are approached. The number of cells, points, and other specifics used in three different meshes is presented in Table 1. Mesh M3 is used for all calculations with viscoelastic fluids. The difference in mesh tessellation employed for cases resulting in transient flows of EVP fluids is required by the flow field characteristics near the contraction plane, where yield surfaces evolve over time. Additionally, there is no need for refinement in the proximity of the re-entrant and salient corners, due to the absence of lip or corner vortices in the EVP fluids. The specifics of the mesh convergence for EVP fluids will be elaborated in the relevant section (Section 4.1).

| Mesh | Number of cells | Number of points | Number of cells in the x / y direction in the wide channel | Number of cells in the x / y direction in the narrow channel |
|---|---|---|---|---|
| M1 | 12000 | 24508 | 66 / 151 | 90 / 45 |
| M2 | 51200 | 103462 | 220 / 210 | 100 / 50 |
| M3 | 99200 | 199962 | 260 / 320 | 400 / 80 |

Table 1- Characteristics of the meshes used in the contraction flow for VE fluid.

## 2.2 Method of numerical solution

To solve the above equations, we have used the finite volume method (FVM), as applied in the open-source package OpenFOAM coupled with the toolbox RheoTool [6] and the library *viscoelasticFluidFoam* [26]. The latter has introduced certain modifications in RheoTool, that focus on the stress-velocity coupling, a sparse matrix solver, and a particular discretization scheme for the convective terms to enhance the stability of the solver. Additional modifications to the FVM have been proposed recently [27-29], but not implemented yet in OpenFOAM.

At each time step, the governing equations are solved iteratively in the following sequence: momentum, continuity (pressure correction), and finally the constitutive equation. A transient simulation with appropriate initial condition is performed using the first-order Euler scheme until a steady state or a periodic solution is achieved or the final set time is reached [28]. For the pressure-velocity coupling, especially for the high-Weissenberg number problem (HWNP), the Semi-Implicit Method for Pressure-Linked Equations (SIMPLE) algorithm is used. For the stress-velocity coupling, the pressure gradient and polymeric extra-stress on the neighbouring cell-faces are used to correct the velocity value at cell centers [30]. The Convergent and Universally Bounded Interpolation Scheme for the Treatment of Advection (CUBISTA) scheme [31] is used for the discretization of the convective terms, which arise in both the constitutive and momentum equations, because it improves the stability and accuracy of the numerical simulation [6]. It is noteworthy that a fairly small solvent viscosity is introduced in all simulations in our study, because it is necessary to retain the ellipticity of the momentum equation and stabilize the solution.

To address the high Weissenberg number problem (HWNP), the log-conformation formulation has been adopted, which takes advantage of the positive definiteness of the conformation tensor. The mathematical details of each step are elaborated in [32] and the process is employed in RheoTool. One of the main concerns in solving the continuity equation in the FVM is the very weak diagonal dominance of the coefficient matrix. Among the various iterative solvers to tackle this issue [26], the velocity and stress fields are solved with a preconditioned conjugate gradient solver (PBiCG) with the diagonal-based incomplete LU (DILU) preconditioner. For the pressure field, a preconditioned conjugate gradient solver (PCG) with the diagonal based incomplete Cholesky preconditioner (DIC) is used. Other details of the RheoFoam solver are elaborated extensively in [6]. The time-step is $10^{-4}$ and the maximum Courant number ($Co = \frac{u\Delta t}{\Delta x}$) is set at 0.1 in all simulations. The maximum relative tolerance for convergence of iterations was $10^{-6}$, for the velocity, pressure, and stress fields.

## 3  Validation of the model and solution method for Viscoelastic fluids

Prior to delving into the EVP results, it is imperative to validate the implementation and accuracy of our numerical tool by comparing the viscoelastic fluid flow results with recent reports in the literature for the Oldroyd-B fluid model, because it produces the most sensitive results. First, we examine the velocity profile to ensure that the fully developed condition has been attained well before the contraction plane. Then, we compare our predicted size of the lip and corner vortices in contraction flow, with those in the literature.

### 3.1  Fully developed flow conditions

Achieving fully developed conditions is imperative in order to eliminate any effect from the inlet conditions and, hence, isolate the influence of the contraction on the flow. Fig. 2 illustrates the velocity profile at half the distance between the entrance and the contraction plane, $\frac{x}{w} = -50$. The profile is fully developed, parabolic, and closely follows earlier experimental and numerical reports [33].

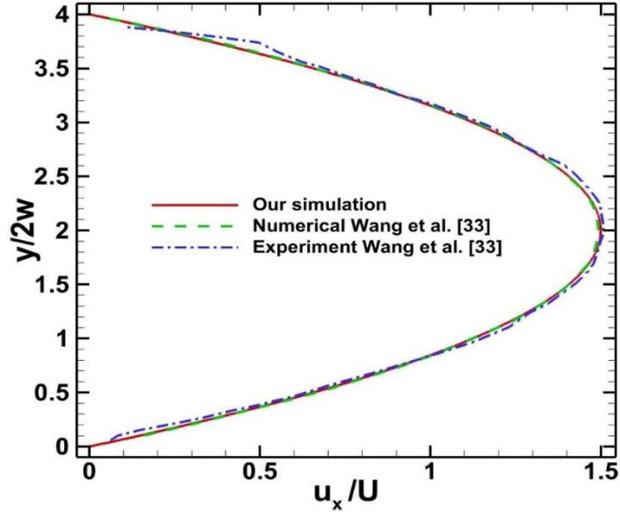

Figure 2. Comparison of the numerical velocity profile with the experimental measurements for viscoelastic fluid at $\frac{x}{w} = -50$ for contraction flow ($Wi$=1.7, $Re$=0.01).

## 3.2 Lip and corner vortices

According to previous studies [6, 34, 35], two vortices appear at the re-entrant and concave corners of this geometry when the Oldroyd-B fluid model is used. The size of these vortices is most sensitive to material parameters and flow conditions, and, hence, it is most suitable to validate numerical simulations. We examine them for a wide range of $Wi$, ($0 \leq Wi \leq 12$) and different meshes to ascertain the accuracy and convergence of our predictions and compare their size with previous studies, which have used FEM or FVM [6, 34-37].

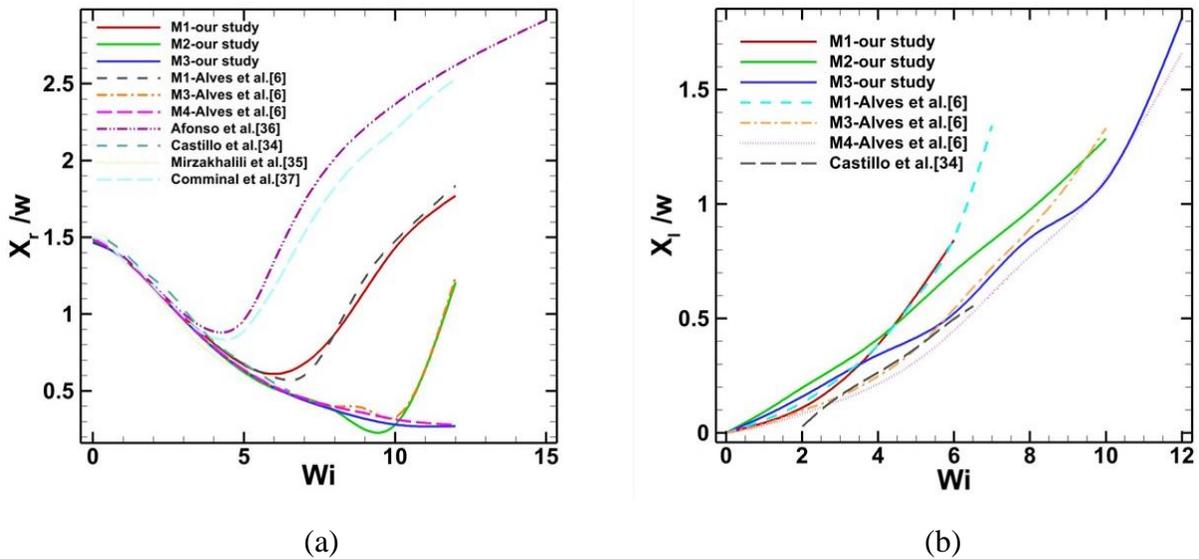

(a) (b)

Figure 3. Comparison with literature results of the dimensionless length of (a) the corner vortex, $X_r/w$ and (b) the lip vortex $X_l/w$ versus the Weissenberg number using our 3 VE meshes.

In Fig. 3(a) the length of the corner vortex (defined in Fig. 1) versus the Weissenberg number is depicted. As $Wi$ increases this length decreases, until a critical value of $Wi$ beyond which it starts increasing rather sharply. Apparently, this critical value increases with mesh refinement. The approximate critical $Wi$ for our mesh M1 and M2 is 6 and 9.5, respectively. Our predictions with the finest mesh, M3, have not reached such a critical value but coincide with the predictions of the other two meshes up to their critical values. A good agreement can be seen between the three different meshes and the results of previous studies. Therefore, special care is needed to capture the rather slow flow in the corner, because it strongly depends on the quality of the mesh there [6]. This is the reason we have locally refined the mesh, as seen in Fig. 1.

In Fig. 3(b) the length of the lip vortex (defined in Fig. 1) versus the Weissenberg number is illustrated. It increases with $Wi$, and decreases with mesh refinement. The same influence of the mesh refinement on the lip vortex size arising in extrusion from a die has been reported in [38]. As was mentioned previously the lip vortex size is the most sensitive parameter, and, hence, it is strongly affected by mesh refinement. Castillo and Codina [34] reported the lip vortex size up to $Wi = 7$, which compares favorably with the prediction of mesh M3 in our study. In Fig. 3(b), the value of the lip vortex for mesh M1 and M2 is not reported after $Wi = 6$ and 10, respectively. The reason is that at these values of $Wi$ the corner and lip vortex have merged and only the corner vortex size is reported. Since at the singular points of the contraction flow, the stress component values grow unbounded, it is very difficult to achieve completely mesh-independent results, especially at high $Wi$ numbers [7], without continuous local mesh refinement (Fig. 1), which sharply increases the computational cost [6]. On the other hand, we are interested in simulating EVP fluids using the SHB model, which is shear thinning, and, consequently, a lip vortex is not expected to arise.

In Fig. 4 the effects of $Wi$ on the streamlines and the lip and corner vortices can be seen. As demonstrated in Fig. 3, increasing $Wi$ leads to an increase of the length of the lip and a decrease of the length of the corner vortex. When $Wi$ and fluid elasticity increase, the streamlines in the wide channel are seen to remain nearly straight further downstream or even diverge as they approach the wall, which is vertical to the main flow direction. This necessitates an increasingly more abrupt redirection of the fluid towards the entrance of the narrow channel, which translates into a decreasing local radius of curvature near both the salient and re-entrant corner. It has been established in [39, 40] that the combined effect of increasing fluid elasticity and decreasing radius of curvature of the streamlines may lead to a purely elastic instability. Indeed, instability has been reported in [36] for this flow but for $Wi > 13$, where the flow first loses its plane of symmetry and then becomes transient. We have reported steady flow with Weissenberg as high as $Wi = 12$ in Fig. 4, so our predictions agree with those in [36].

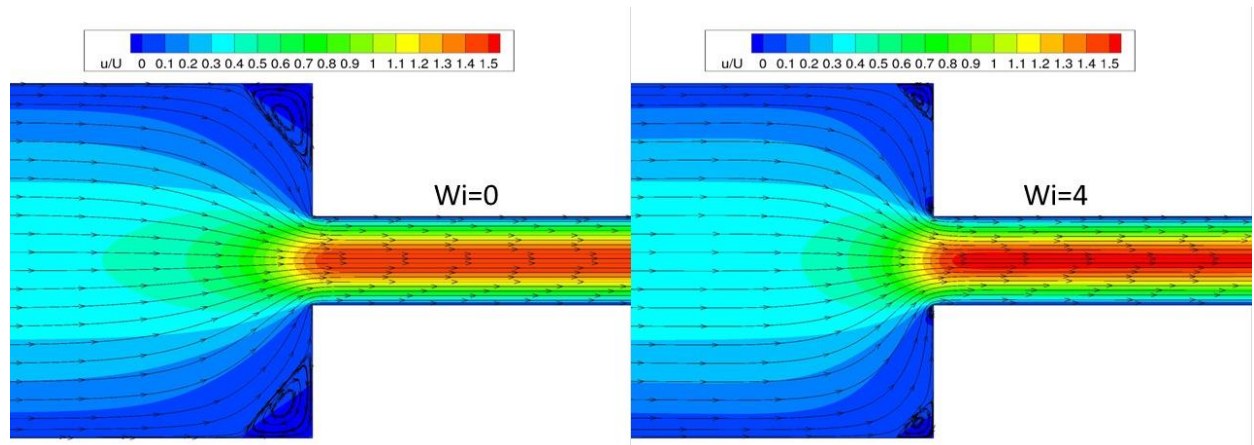

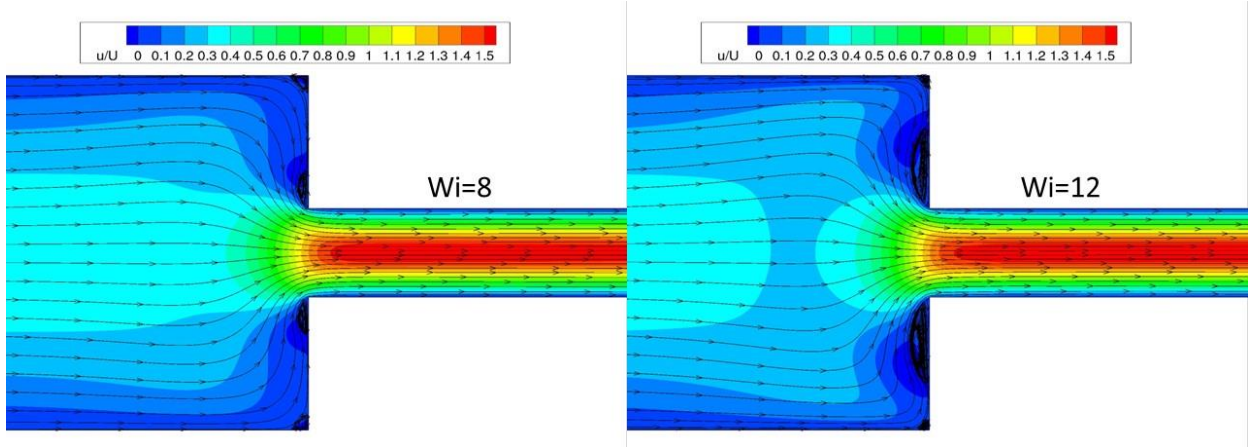

Figure 4. Velocity contours and the changes of lip and corner vortices size for 4 values of $Wi$ for an Oldroyd-B fluid model.

## 4 Results and discussion

### 4.1 Flow of an Elastoviscoplastic fluid in a 4:1 planar contraction

The EVP material behaves as a hyperelastic solid before reaching its yield stress, whereas it behaves as a viscoelastic and shear thinning liquid above this value. First, we compare our predictions for three rheologically characterized Carbopol gels. Subsequently, we investigate the effect of the four SHB parameters on the results and particularly when the elastic modulus is above, or the yield stress is below a critical value and lastly when these critical values are surpassed.

Tables 2 & 3 present the rheological parameters measured for these three EVP fluids and the corresponding non-dimensional numbers arising in the contraction flow. We have used as a typical shear rate, $\frac{U}{w} = 0.4 \ 1/s$ which implies that for $w = 0.001 \ m$, the exit velocity is $U = 0.0004 \ m/s$, as in the paper by Alves et al. [6]. The values of the Reynolds and Weissenberg numbers are less than 0.05, whereas the Bingham number is $O(1)$, so it dominates the fluid behavior. In this study, we introduce a solvent viscosity of $\eta_s = 0.01 \ Pa \ s$ - in line with the water viscosity. As stated earlier, a nonzero $\eta_s$ is required by OpenFOAM. This value of $\eta_s$ results in a fairly small viscosity ratio $\beta = O(10^{-3})$, because solvent viscosity was not even considered in the rheological characterization of these fluids [18, 19, 24].

| Rheological Parameters | $n$ | $\tau_0 \ (Pa)$ | $k \ (Pas^n)$ | $G \ (Pa)$ |
|---|---|---|---|---|
| Lopez et al. [18] (0.09%) | 0.45 | 3.08 | 1.75 | 20.55 |
| Lopez et al. [18] (0.1%) | 0.46 | 4.71 | 1.81 | 40.42 |
| Pourzahedi et al. [19] (0.1%) | 0.42 | 4.61 | 3.4 | 41.8 |

Table 2- Rheological parameters of Carbopol 0.09% and 0.1% concentrations taken from [18, 19, 24].

| Source of Experimental Data | $Re = \dfrac{\rho U w}{\eta_s + k(\frac{W}{U})^{1-n}}$ | $Wi = \dfrac{k(\frac{U}{W})^n}{G}$ | $Bn = \dfrac{\tau_0 w}{(\eta_s + k(\frac{W}{U})^{1-n})U}$ | $\beta = \dfrac{\eta_s}{\eta_s + k(\frac{W}{U})^{1-n}}$ | $\varepsilon_y = \dfrac{\tau_0}{G}$ |
|---|---|---|---|---|---|
| Lopez et al. [18] (0.09%) | 0.00013 | 0.0563 | 2.649 | 0.00344 | 0.1498 |
| Lopez et al. [18] (0.1%) | 0.00013 | 0.0293 | 3.953 | 0.00335 | 0.1165 |
| Pourzahedi et al. [19] (0.1%) | 0.000069 | 0.0553 | 1.988 | 0.00172 | 0.1102 |

Table 3- Non-dimensional numbers for the contraction flow based on the rheological parameters of Carbopol 0.09% and 0.1% concentrations taken from [18, 19].

Firstly, we conducted a mesh convergence study focused on the contraction flow for EVP materials. Four mesh configurations were employed for the contraction flow, and their details are outlined in Table 4. Each subsequent mesh was generated primarily by locally refining the previous one, so we avoided increasing the computational cost excessively. No previous studies exist to compare with. For all comparisons we use the Carbopol 0.1% material properties obtained via linear regression on the data reported by Lopez et al. [18].

| Mesh | Number of cells | Number of points | Number of cells in the x / y direction in the wide channel | Number of cells in the x / y direction in the narrow channel |
|---|---|---|---|---|
| M1 | 79600 | 160542 | 260 / 260 | 300 / 80 |
| M2 | 99200 | 199962 | 260 / 320 | 400 / 80 |
| M3 | 120000 | 241832 | 275 / 320 | 640 / 100 |
| M4 | 142000 | 285992 | 275 / 400 | 640 / 100 |

Table 4- Characteristics of the meshes used in contraction flow for EVP fluid.

In Fig. 5, the profiles of the dimensionless velocity ($u_x/U$) and normal stress component ($\tau_{xx}/\tau_o$) along the $x$-axis at $y$=0 are depicted. Note that the stress has been normalized by the yield stress in this figure and hereafter. The primary velocity exhibits an overshoot at the entrance of the narrow channel, which is caused by a combination of the converging flow and fluid elasticity.

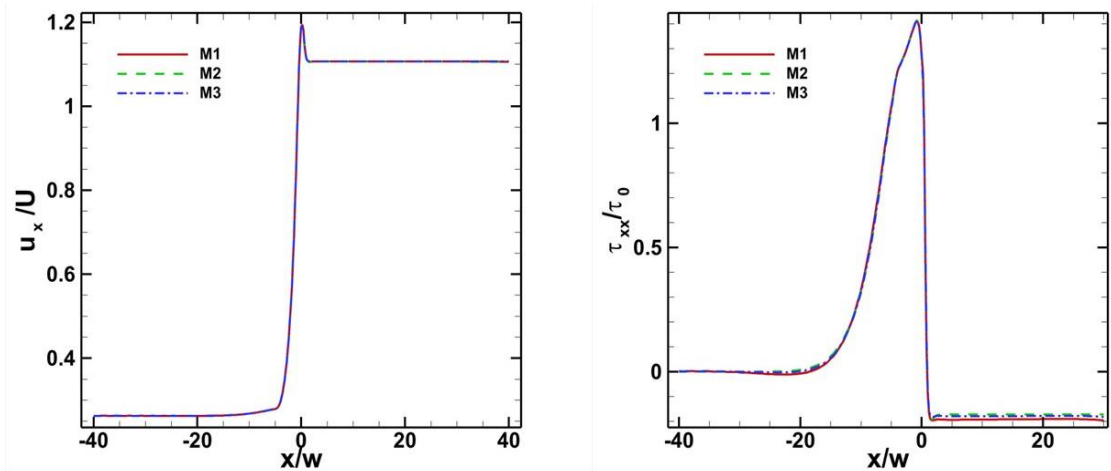

(a)                        (b)

Figure 5. (a) dimensionless velocity ($\frac{u_x}{U}$) and (b) normal stress component ($\frac{\tau_{xx}}{\tau_0}$) along the $x$-axis at $y=0$.

The normal stress component in the main flow direction experiences a sharp increase starting well before the entrance of the narrow channel, which is maximized just before it, and is relaxed to negative values within the narrow channel. These negative values arise from the significant compression of the material caused by the change in cross-section of the channel. As in viscoelastic flows, in the vicinity of the re-entrant corners, there is a rapid increase in all normal stress components, followed by a swift decay. Both the velocity and stress overshoots are accurately captured using all three meshes, so in all simulations involving EVP materials in this geometry and leading to a steady state, the intermediate mesh, M2, is employed, whereas mesh M4 is used for simulations leading to extended transients or chaotic flows. Additional mesh convergence tests when the flow either finally becomes steady or remains transient are provided in Appendix A.

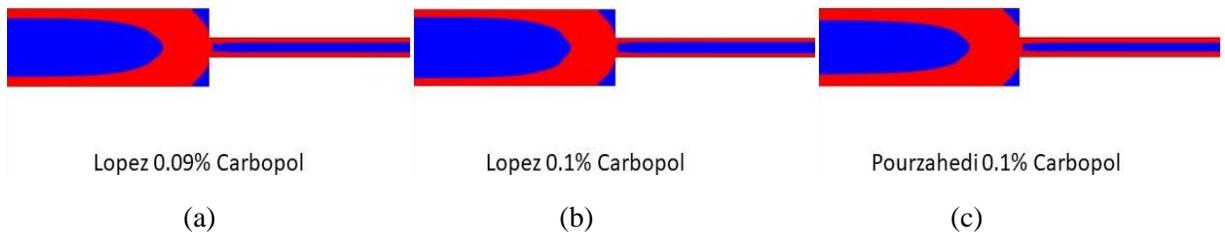

Lopez 0.09% Carbopol          Lopez 0.1% Carbopol          Pourzahedi 0.1% Carbopol

(a)                  (b)                  (c)

Figure 6. Illustration of the yielded (red) and unyielded (blue) along the $x$-axis ($-22 \leq \frac{x}{w} \leq 22$) for the three sets of Carbopol solutions listed in Table 2, [18, 19].

In Fig. 6 the yielded and unyielded zones of the EVP material are shown for Carbopol 0.09 and 0.1% concentrations [18, 19]. There are two unyielded zones located around the midplane, one in each channel and away from the contraction area and two additional ones in the concave corners. Despite the very close concentrations and rheological parameters, one can readily distinguish differences in the yield surfaces between the three fluids. For example, the yielded area next to the wall is narrower and the front of the yield surface is sharper in the Lopez 0.1% Carbopol. Overall, these areas are not too different from the corresponding ones in Bingham fluids [17].

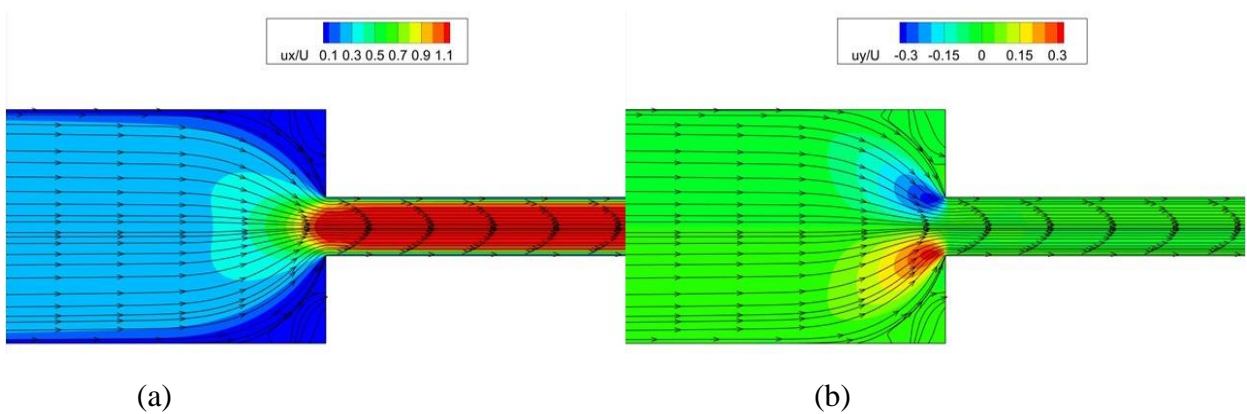

(a)                        (b)

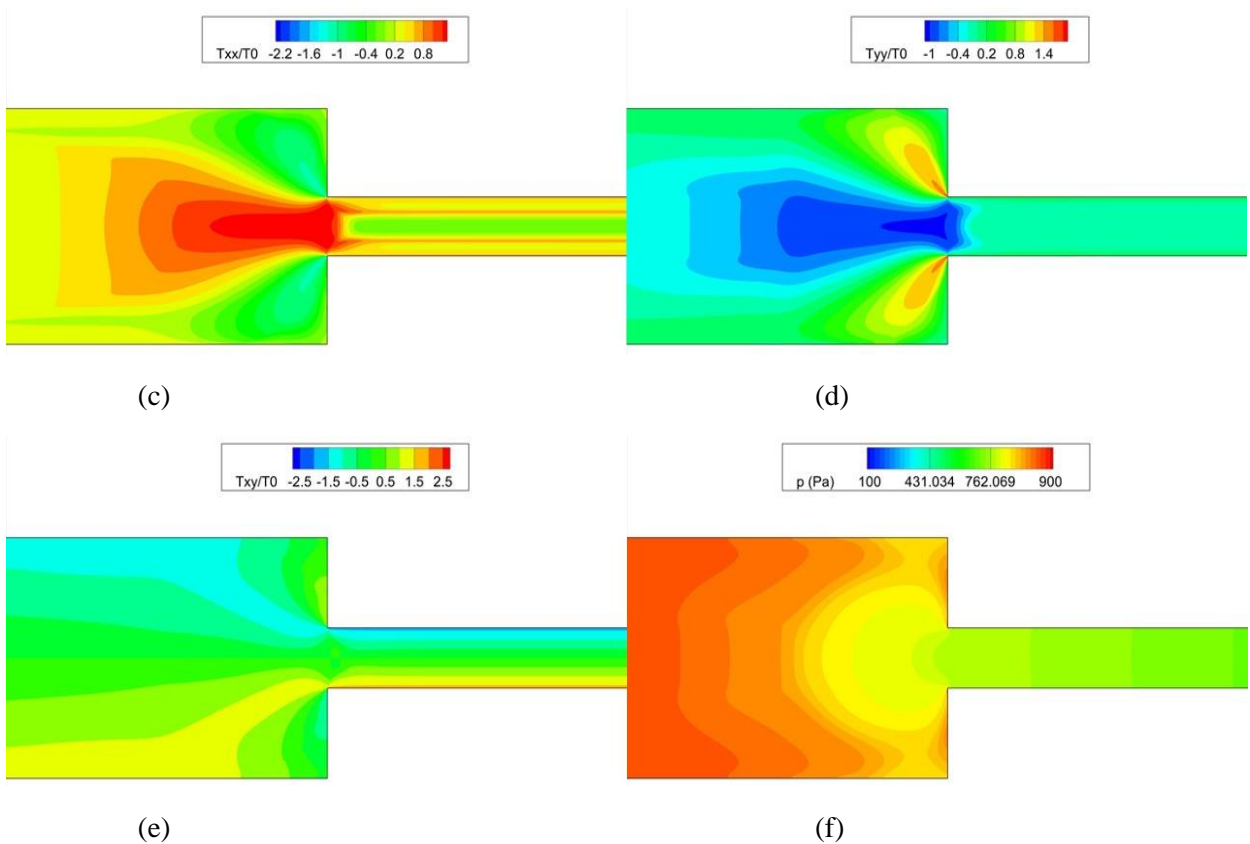

Figure 7. Fields of $(a) \frac{u_x}{U}, (b) \frac{u_y}{U}, (c) \frac{\tau_{xx}}{\tau_0}, (d) \frac{\tau_{yy}}{\tau_0}, (e) \frac{\tau_{xy}}{\tau_0}, (f) p$ in the contraction geometry ($-11 \leq \frac{x}{w} \leq 11$) for the 0.1% Carbopol solution listed in Table 2 by Lopez et al. [18]. Along with the velocity fields, streamlines are also presented in (a) and (b)

Fig. 7 illustrates the fields of velocities, stresses, and pressure within the wide and narrow channel near the contraction. The primary velocity is uniform in most of the cross-section of both the wide and narrow channels and exhibits strong shearing very close to the long walls and a sudden increase as the fluid enters the narrow channel. The secondary velocity retains a very small magnitude throughout the geometry, except near re-entrant corners where it increases abruptly, to negative values in the upper half and positive ones in the lower one, demonstrating that the fluid is forced to enter the contraction. In both concave corners, there is slow recirculation, indicating that the material is below the yielding limit there. The velocity field is in agreement with the locations of yielding/unyielding material seen in the middle of figure 6. The normal stress component, $\tau_{xx}$, is positive around the centerplane, and keeps increasing as the entrance of the narrow channel is approached, indicating fluid extension in the primary flow direction. On the contrary, around the same plane $\tau_{yy}$ is negative and keeps decreasing up until the entrance of the narrow channel, indicating fluid compression in the $y$-direction. Its magnitude is similar to that of $\tau_{xx}$. In a narrow section between the re-entrant and salient corners, the signs of the normal stress components are reversed, but their magnitudes remain similar, as the fluid is forced to approach the contraction. Both normal stress components are nearly constant in most of the narrow channel and around the plane of the symmetry, again concurring with the existence of the unyielded area there. In the part of the narrow channel closer to the walls $\tau_{xx}$ becomes somewhat larger than $\tau_{yy}$. As expected, the shear stress is considerably larger in magnitude near the walls, with the opposite signs and zero at the centerplane. Finally, the pressure decreases in the flow direction. Its variation is smaller within the narrow channel as opposed to the wider one. It is noteworthy that the iso-surfaces of all stress components and pressure are smooth, which testifies once more to the accuracy of the numerical results.

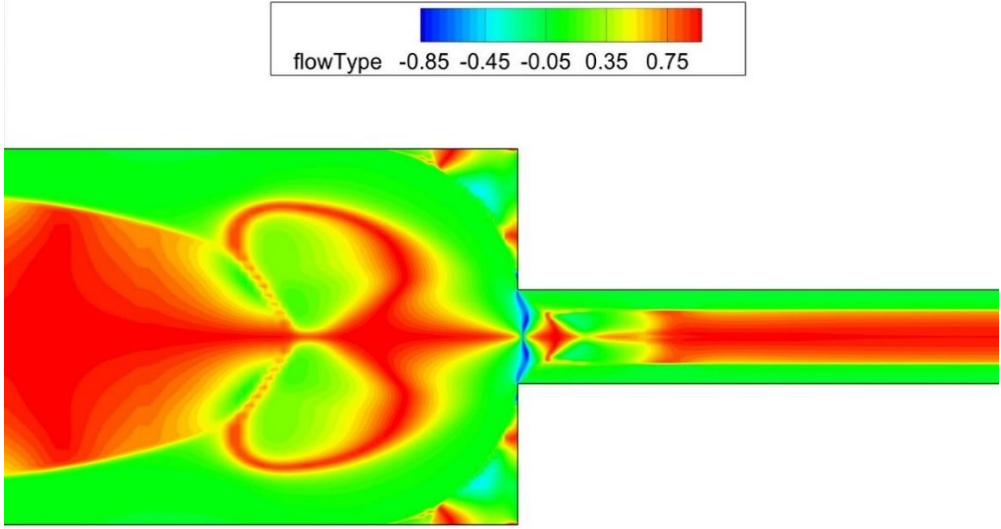

Figure 8. Flow type parameter $Q$ distribution for Lopez Carbopol 0.1%.

The type of flow dominating in different parts of the geometry can be determined by the flow-type parameter [41, 42], defined as:

$$Q = \frac{|D|-|\Omega|}{|D|+|\Omega|}, \boldsymbol{D} = \frac{1}{2}(\nabla \boldsymbol{u} + (\nabla \boldsymbol{u})^T), \boldsymbol{\Omega} = \frac{1}{2}(\nabla \boldsymbol{u} - (\nabla \boldsymbol{u})^T), |D| = \sqrt{2(\boldsymbol{D}:\boldsymbol{D})}, \text{ and } |\Omega| = \sqrt{2(\boldsymbol{\Omega}:\boldsymbol{\Omega})}. \quad (10)$$

In the above expressions $\boldsymbol{D}$ is the rate of strain tensor and $\boldsymbol{\Omega}$ is the vorticity tensor. When $Q \sim -1$, the flow is dominated by rotation, when $Q \sim 0$, it is dominated by shear, and when $Q \sim 1$, it is dominated by extension. Fig. 8 shows that extensional flow prevails at the center of both the wide and the narrow channels, whereas close to the channel walls, shear flow dominates. Near the entrance of the narrow channel, a small area of predominantly rotational flow is observed. Following the central unyielded region in the wide channel, where extension dominates, two circular regions appear dominated by shear flow. Shear also dominates upstream from the wall, which is vertical to the main flow direction. These shear areas are generated by the motion of yielded material from the thin shearing film close to the walls in the wide channel to its center. This deflection is driven by the contraction in the geometry and elevates the shear contribution close to the main unyielded region.

In the following, we will present only the variations of the yield surface caused by changing the rheological parameters one at a time of the Carbopol 0.1% [18], which, of course, affect the dimensionless numbers.

### 4.2 Parametric study when a steady solution is reached

In Fig. 9 the effect of the shear thinning exponent, $n$, on the yielded/unyielded regions is depicted. We have restricted the values of the exponent to $n \leq 0.5$. As explained before [43, 44], extension-rate hardening cannot be observed in typical EVP materials, such as Carbopol, emulsions, etc., because of their structure.

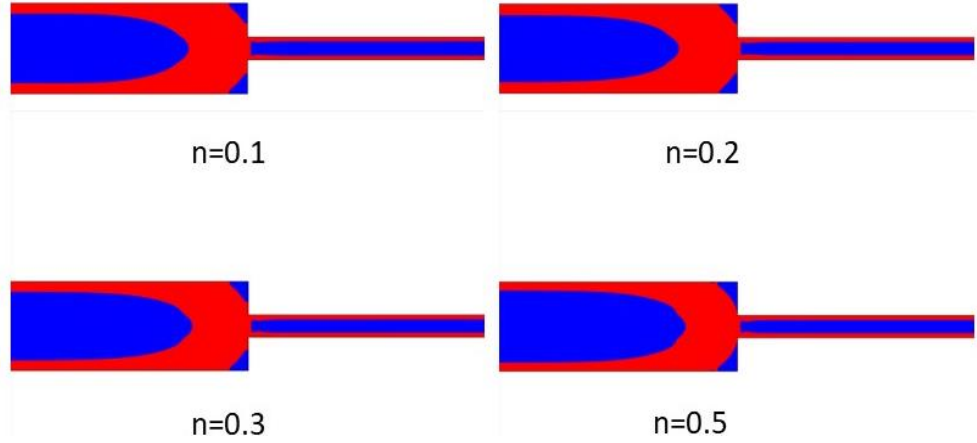

Figure 9. Effect of the Herschel-Bulkley exponent, $n$, on the yielded (red) and unyielded (blue) zones, the rest of the parameters are those of Carbopol 0.1% in [18].

This sets a limit of 0.5 to the Herschel-Bulkley index. Increasing $n$ increases $Re$ (remaining very small) and $Bn$, but decreases $Wi$. In Fig. 9, we observe that increasing $n$ causes the unyielded zone in the wide channel to expand towards the channel walls, while in the narrow channel it remains nearly unchanged. Moreover, increasing $n$ the front of the unyielded area in the wide channel develops an off-centered indentation and the vertical length of the unyielded area at the concave corner increases.

Fig. 10(a) defines certain characteristic lengths of the unyielded/yielded area in the two channels. $Y_1$ is the thickness of the yielded film next to the wall in the wide channel where the flow is fully developed and not affected by the inlet or contraction plane. $Y_2$ is the length of the unyielded material from the salient corner to the entrance to the contraction in the wide channel. $Y_3$ is the thickness of the yielded film in the narrow channel where the flow is fully developed and not affected by the outlet or contraction plane. $X$ is the distance of the tip of the main unyielded region in the wide channel from the center of the coordinate system. Given that $Bn$ is the dominant dimensionless number and it increases with increasing $n$, we observe in Fig. 10(b), as expected, that this increase in $n$ (see definitions in Table 3) leads to a decrease of the yielded thickness well upstream of the contraction, which is linear, and a sharper also linear decrease of the distance of the tip of the unyielded area from the contraction plane, while the yielded thickness well downstream of the contraction remains nearly constant. The width of the unyielded area in the salient corner increases also monotonically. This width changes more abruptly for $n \in (0.3, 0.46)$, because a rather thin unyielded area suddenly develops near the re-entrant corner at the entrance of the narrow channel, which merges with the unyielded area at the concave corner. For $n \in (0.46, 0.5)$ the unyielded area covers almost completely the wall normal to the main flow direction, so $Y_2/w \sim 3$.

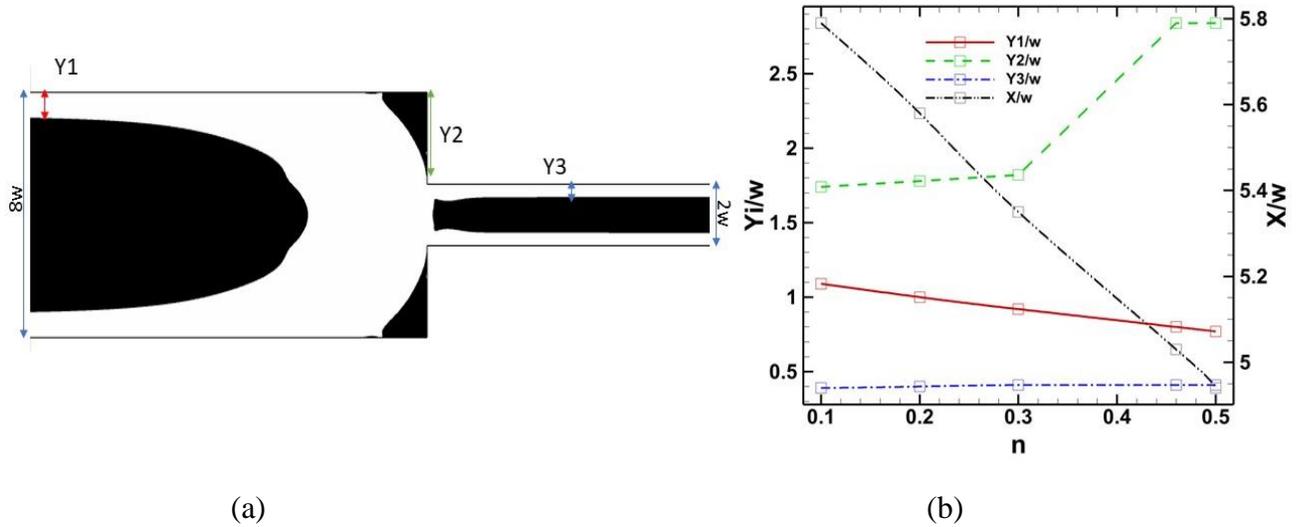

(a) (b)

Figure 10. (a) Definition of characteristic lengths of the yielded area, (b) dependence of these characteristic lengths on the Herschel-Bulkley exponent, $n$.

Fig. 11(a) illustrates the effect of the consistency index, $k$, on the yielded/unyielded regions. The value of the consistency constant has a different effect on the three non-dimensional numbers. According to Table 3, when $k$ increases, $Re$ and $Bn$ decrease, while $Wi$ increases (remaining less than one). Therefore, the primary effect is on $Bn$ and the highest value of $k$ results in the smallest unyielded regions. This is quantified in Fig. 11(b), where we observe that increasing $k$ leads to an increase of the yielded thickness well upstream and well downstream from the contraction, and a sharper increase of the distance of the tip of the unyielded area from the contraction plane. All three characteristic lengths present a parabolic dependence on $k$. The width of the unyielded area in the salient corner decreases monotonically and more abruptly when $k \in (1.81, 5)$, because the rather thin unyielded area closer to the entrance of the narrow channel suddenly vanishes (opposite to what happened when the effect of increasing $n$ was examined).

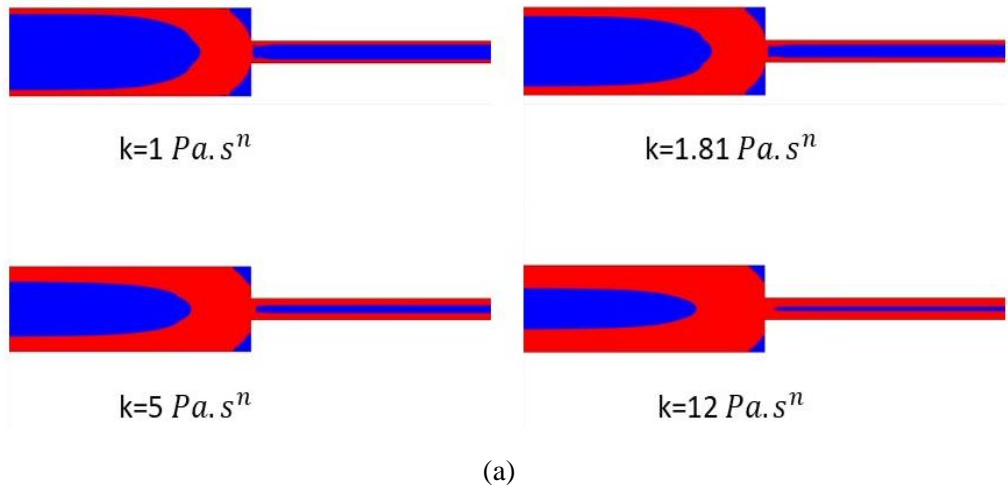

(a)

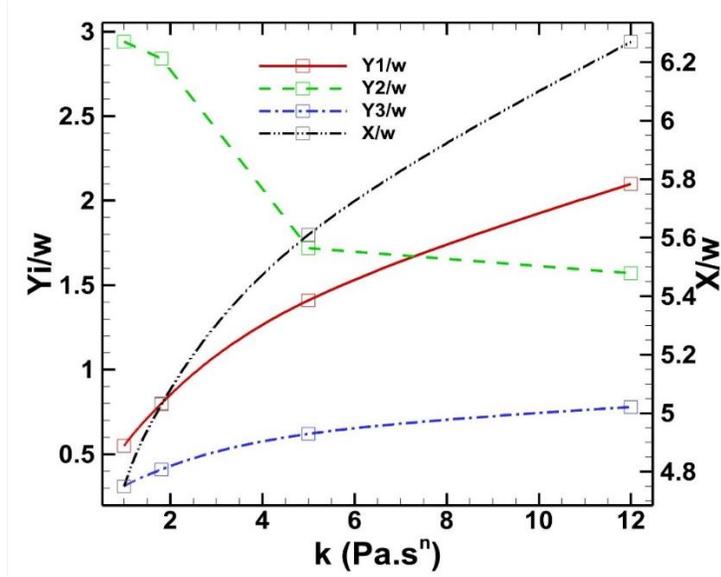

(b)

Figure 11. (a) Effect of $k$ on the yielded (red) and unyielded (blue) zones, and (b) Dependence of the characteristic lengths on $k$. The rest of the parameters are those of Carbopol 0.1% in [18].

The value of the yield stress affects proportionately $Bn$, but no other dimensionless number. Its increase enlarges the unyielded zones in the channels as seen in Fig. 12(a). Furthermore, the dead zones at the concave corners increase primarily towards the re-entrant corners. All these changes are quantitatively shown in Fig. 12(b). Here, we observe that increasing $\tau_0$ leads to a nearly hyperbolic decrease of the yielded thickness well upstream of and well downstream from the contraction. Similarly, the distance of the tip of the unyielded area from the contraction plane decreases monotonically. On the contrary, the width of the unyielded area in the salient corner increases rapidly up to $Bn \approx 3$, when it covers the entire length of the wall normal to the main flow, and, of course, remains nearly constant thereafter.

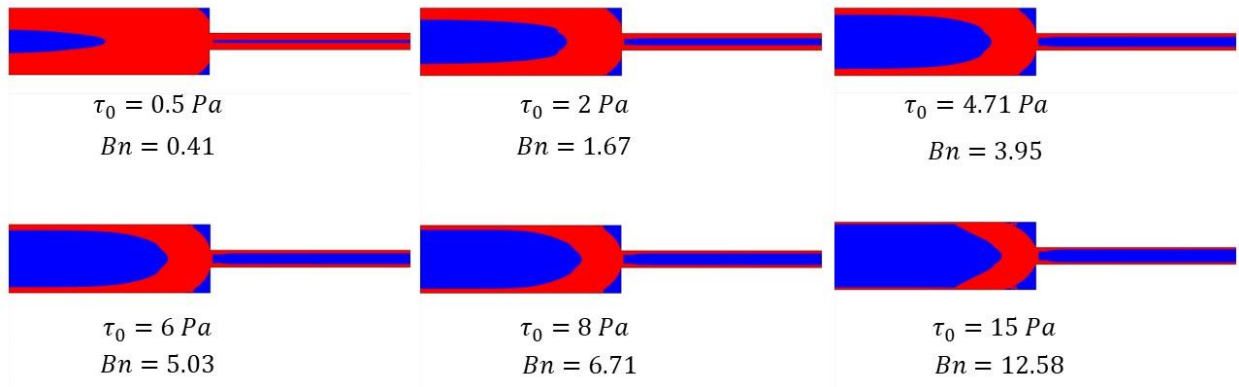

(a)

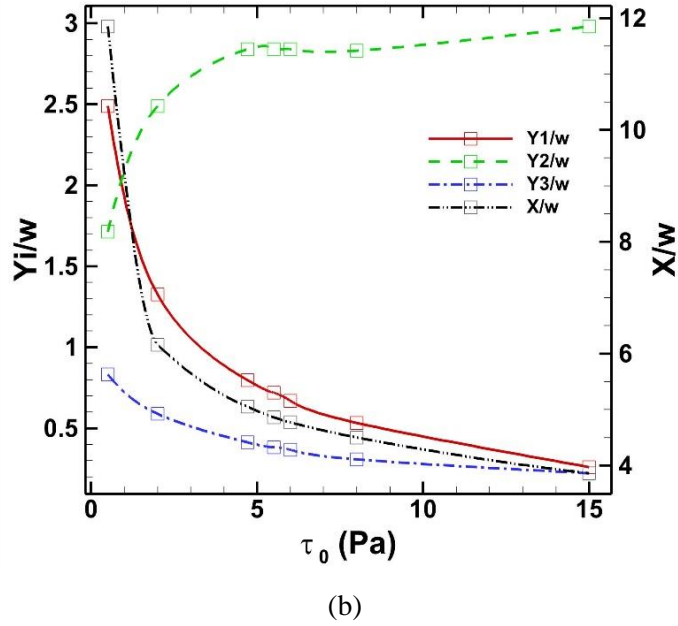

(b)

Figure 12. (a) Effect of $\tau_0$, on the yielded (red) and unyielded (blue) zones, (b) Dependence of the characteristic lengths on $\tau_0$. The rest of the parameters are those of Carbopol 0.1% in [18].

Fig. 13 (a) depicts the effect of the elastic modulus, $G$, on the yielded/unyielded regions. The elastic modulus is inversely proportional to $Wi$, but does not affect any other dimensionless number. As $G$ decreases, the material becomes more elastic. This allows the material to deform more prior to yielding, and hence, the unyielded area in the wide channel to expand. For the same reason, the tip of the unyielded zone becomes sharper over a longer distance and translates in the direction of flow. The same increase of the unyielded areas with increasing elasticity in an EVP material has been predicted in model porous media [45] and in shear flow of an EVP droplet in a Newtonian medium [46]. Increasing $G$ leaves nearly unaffected the unyielded area in the narrow channel and around the salient corner. These observations are quantified in Fig. 13 (b). The reason for this is that we are only decreasing $G$ keeping the values of the other three parameters to their values for the base material with which unyielded material covers completely the distance from the salient to the re-entrant corner, so $Y_2 \sim 3$. Unyielded material exists also upstream from this part of the wall. Hence, the material is only slightly deformed and flows extremely slowly in this area (the same holds even when material plasticity is absent) and is not affected by the elastic stress. Increasing material elasticity in the range $5\ Pa \leq G \leq 50\ Pa$, where a steady state is reached, cannot affect this region (again the same holds in the absence of plasticity). Material elasticity must be decreased further (below a critical value) to $G \leq 4\ Pa$ to make the flow transient and affect this region. The transient flow is discussed in the next section of the paper. In the region far downstream from the contraction plane and when $5\ Pa \leq G \leq 50\ Pa$ the flow is 1D again with all velocity and stress components fully developed. In this region varying $G$ or any other rheological parameter ($n, k, and\ \tau_0$) leaves only slightly affected, if at all, $Y_3$, in contrast to the readily observable effect on $Y_1$, the corresponding length in the wide channel.

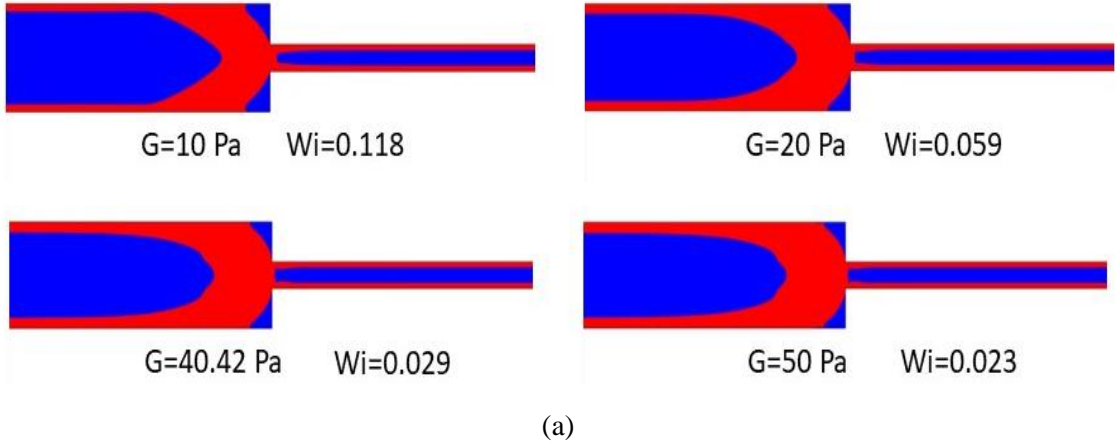

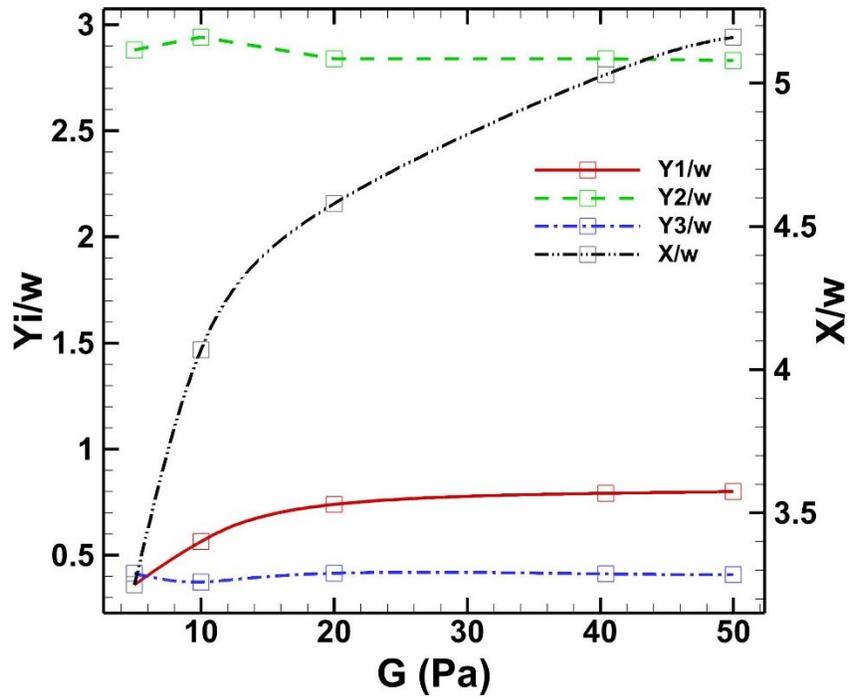

Figure 13. (a) Effect of $G$ on the yielded (red) and unyielded (blue) zones, and (b) Dependence of characteristic lengths on $G$. The rest of the parameters are those of Carbopol 0.1% in [18].

### 4.3 Oscillatory or transient dynamics when $G$ decreases below a critical value

When we extended our study to even lower values of the elastic modulus or larger values of the yield stress, we observed that the flow remained transient. This was quite surprising for a yield-stress fluid in creeping flow, especially under increasing $\tau_o$. So, we consider it important to dedicate a separate section to this phenomenon. Since the resulting velocity and stress fields are quite more complicated, the most refined mesh, M4, given in Table 4 was used in all these simulations. All simulations were carried out until $t = 200\ s$. First, we discuss the effect of decreasing $G$.

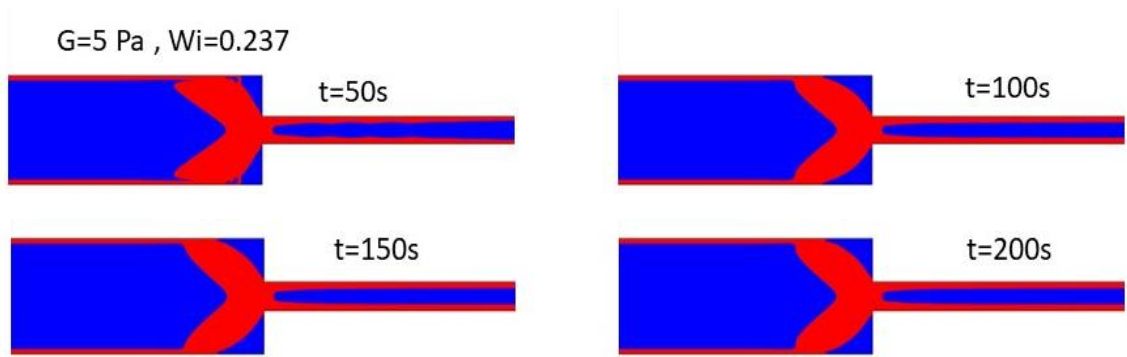

Figure 14. Evolution of the yielded (red) and unyielded (blue) zones for $G = 5\ Pa$ ($Wi = 0.237$), while the rest of the parameters are those of the 0.1% Carbopol solution of Lopez et al. (Table 2). The yield strain of this case is $\varepsilon_y = 0.942$.

Fig. 14 illustrates the evolution of both yielded/unyielded areas for $G = 5\ Pa$, at the four indicated time-instants. A more complete picture of this and the rest of the transient results is given via the videos ( https://veed.io/view/a6b46474-501c-41a9-973d-8051b3aece73 ) included in the Supplementary Material (SM). From the snapshots we may observe that at $t = 100\ s$ the flow is still evolving. At this instant, the unyielded area before the contraction entrance is still somewhat larger than in the following two snapshots, where it seems to have stabilized. Moreover, at $t = 50\ s$ one can clearly observe a wavy yield surface inside the narrow channel. In the corresponding video traveling waves may be seen on the yield surface that slowly damp out. When the steady flow is established, the yielded material well before the contraction and at all snapshots is much thinner ($Y_1 = 0.35w$) than the one seen for $G = 10\ Pa$ in Fig. 13(a). Since a steady state is eventually reached, these additional datapoints are included in Fig. 13(b). Moreover, the front of the unyielded area around the midplane of the wide channel is more pointed, has a concave instead of convex surface moving away from the midplane and has translated downstream ($X = 3.25w$). Similarly, the front of the unyielded material inside the narrow channel is more pointed and translated downstream. The unyielded area in the salient corner has increased in the upstream direction following the trend observed in Fig. 13(a).

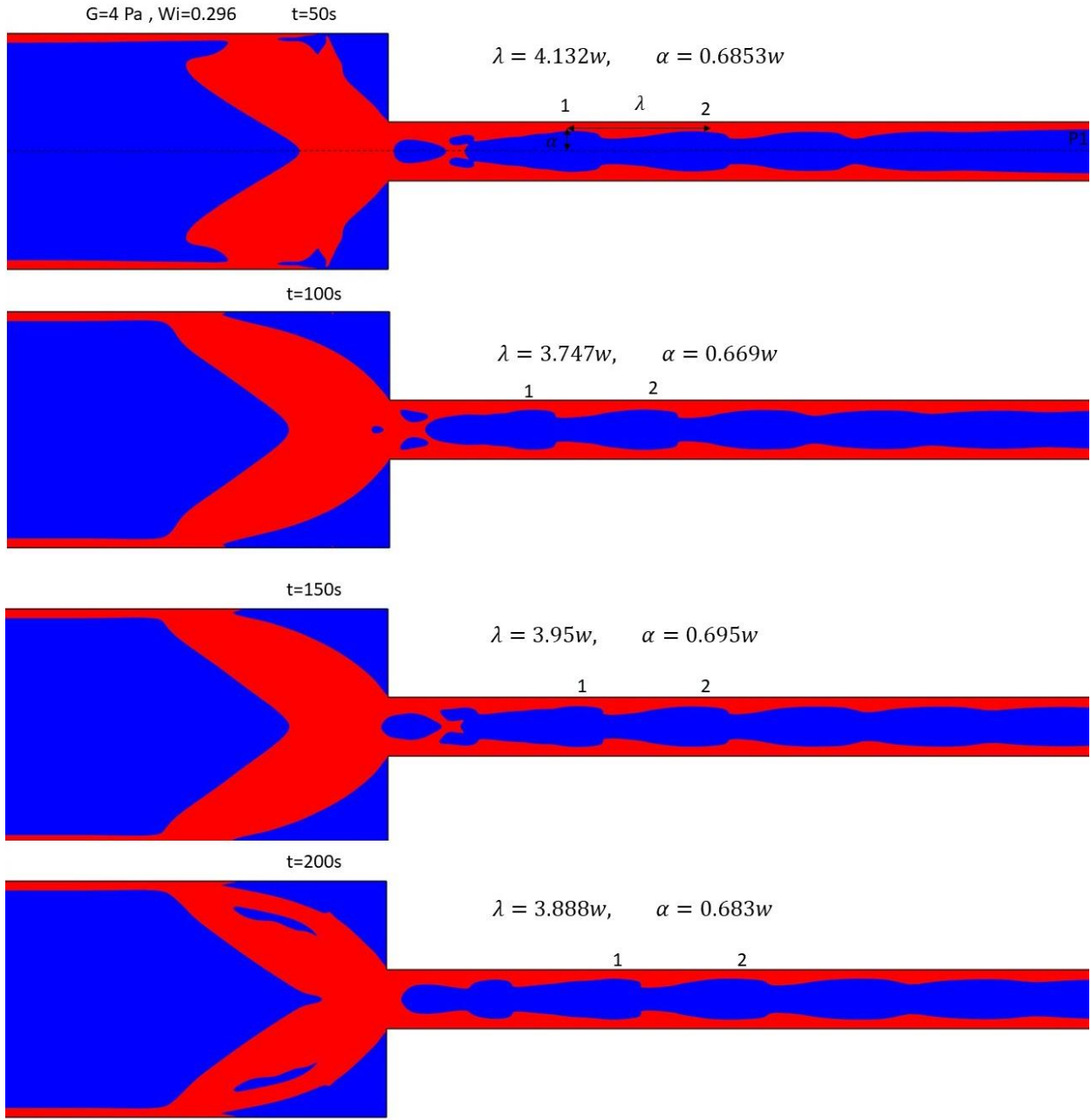

Figure 15. Evolution of the yielded (red) and unyielded (blue) zones for $G = 4\,Pa, (Wi = 0.296)$, while the rest of the parameters are those of the 0.1% Carbopol solution of Lopez et al. (Table 2). The yield strain of this case is $\varepsilon_y = 1.178$.

Fig. 15 illustrates the evolution of both yielded and unyielded areas for $G = 4\,Pa$, at the four indicated time-instants. At $t = 50\,s$ the yield surface even in the wide channel is quite corrugated indicating instability, which is carried into the contraction. In the contraction entrance, the flow varies strongly inducing traveling waves inside the narrow channel again (see the video in SM https://veed.io/view/0e0ffc5a-b9c7-4350-a49d-9db000f87d09). Now their amplitude is larger than in Fig.

14, and their wavelength can be distinguished (see fig. 15) which is decreasing from 4.132w to 3.747w. They seem to damp out downstream, but not in time, as they did in Fig. 14. Even in the wide tube the yield surfaces and, hence the flow, is evolving until the end of our simulations at $t = 200\ s$, because we can observe the formation and disappearance of unyielded islands before the contraction. Moreover, the shape of the front of the yield surface around the midplane changes continuously, becoming more pointed. Only the unyielded area in the salient corners seems to have reached a steady shape, after expanding considerably in the upstream direction. At $t = 200\ s$, the thickness of the yielded film in the wide channel and the distance of the unyielded front to the contraction plane have decreased further to $Y_1 = 0.32w$ and $X = 2.18w$, respectively.

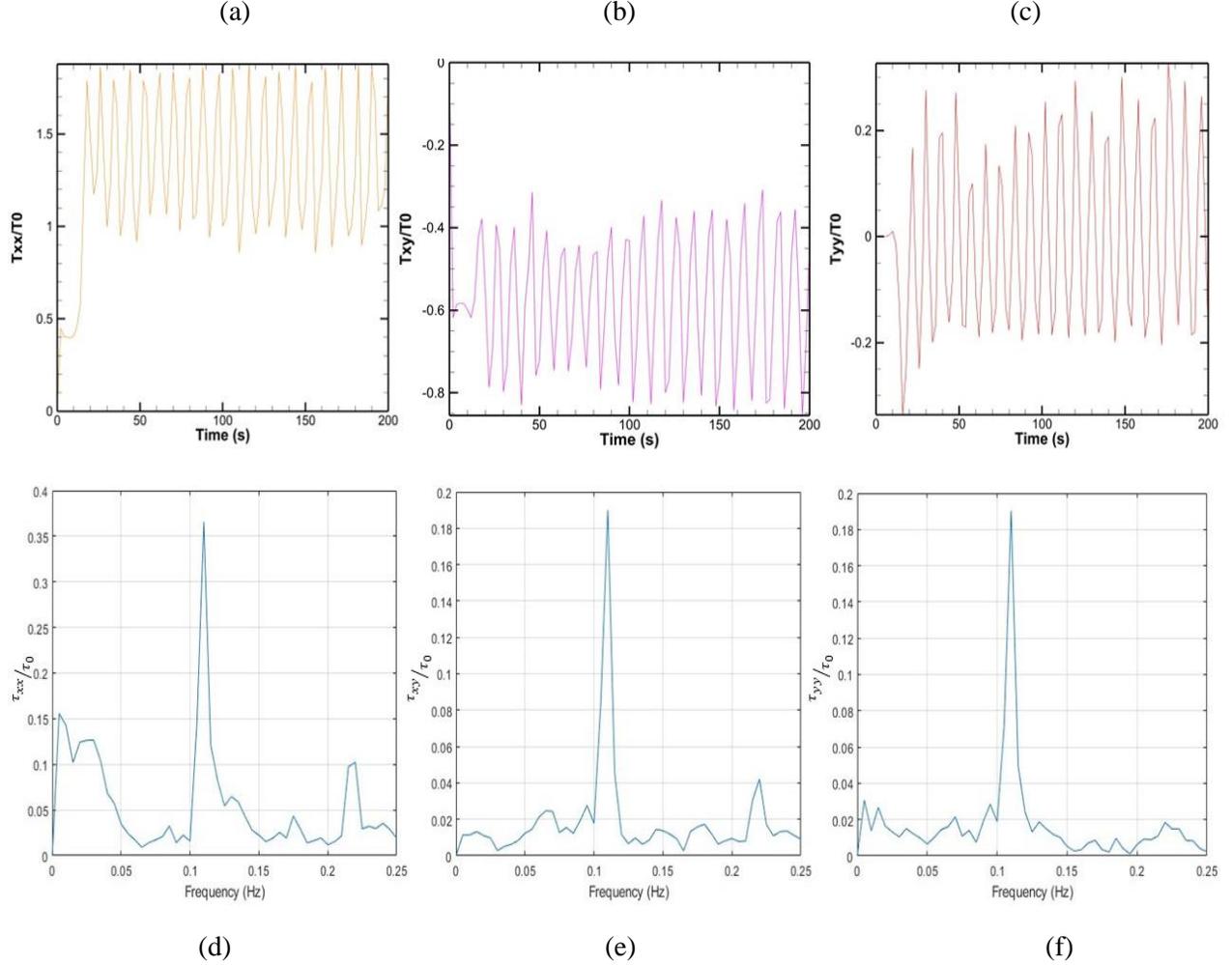

Figure 16. Evolution of dimensionless (a) $\frac{\tau_{xx}}{\tau_0}$ (b) $\frac{\tau_{xy}}{\tau_0}$ (c) $\frac{\tau_{yy}}{\tau_0}$ for the duration of the each simulation for $G = 4\ Pa$ at the probe point $(\frac{x}{w}, \frac{y}{w}) = (5, 0.5)$. The FFT frequency analysis illustrates the dominant and first harmonic frequency for (d) $\frac{\tau_{xx}}{\tau_0}$ (e) $\frac{\tau_{xy}}{\tau_0}$ (f) $\frac{\tau_{yy}}{\tau_0}$.

It is interesting to monitor the evolution of the dependent variables and extract characteristic frequencies from it. We have chosen to examine the stress components, which are the more sensitive variables. Figures 16 (a-c) illustrate the evolution of the three stress components ($\frac{\tau_{xx}}{\tau_0}, \frac{\tau_{xy}}{\tau_0}, \frac{\tau_{yy}}{\tau_0}$) from the beginning of the simulation until the final time ($t = 200s$). The probe point is located near the entrance of the narrow channel, where the primary periodic behavior is evident and away from the plane of symmetry to get a nonzero signal even from the shear stress. All stress components exhibit periodic behavior, as indicated by the corresponding FFT frequency analysis in Figures 16 (d-f). The dominant frequency for $\tau_{xx}$, $\tau_{xy}$ and $\tau_{yy}$ is approximately 0.11 Hz and a weak first harmonic can be seen at ~0.22 Hz. Notably, Fig. 16 (a-c) reveals that $\tau_{xx}$ is higher and always positive than the other stress components, which is another sign of the strong extension of the material at this location. Moreover, $\tau_{xy}$ is always negative, as it should be in this part of the narrow channel, and $\tau_{yy}$ oscillates around zero indicating local periodic contraction and expansion. The velocity evolution and corresponding frequency are presented in Appendix D.

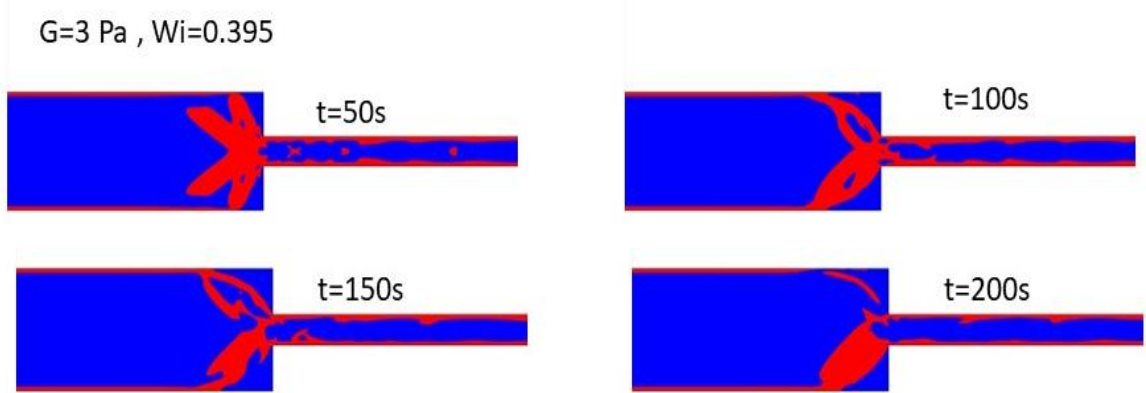

Figure 17. Evolution of the yielded (red) and unyielded (blue) zones for $G = 3\ Pa$ ($Wi = 0.395$), while the rest of the parameters are those of the 0.1% Carbopol solution of Lopez et al. (Table 2). The yield strain of this case is $\varepsilon_y = 1.57$.

Fig. 17 demonstrates the evolution of the yield surfaces in the smallest of the examined values of $G = 3\ Pa$. The yielded and unyielded domains seem to retain as plane of symmetry the midplane, only until $t = 50\ s$, but they are much more corrugated. At this instant we can observe unyielded islands before the contraction and yielded ones in the midplane of the narrow channel. Waves are formed again in the narrow channel, but both their amplitude and wavelength are varying strongly along the $x$-axis. The yield surface in the wide channel is very corrugated and forms a narrow wedge, from which small unyielded islands emanate, see video in SM (https://veed.io/view/0f84901e-6ce8-40c4-8984-d50768677d55). When the island increases in size, it disrupts the flow so much that it makes it asymmetric. Indeed, starting with the next snapshot, $t = 100\ s$, and until the end of the simulations the yielded/unyielded domains have lost their plane of symmetry. This resembles the symmetry breaking reported in [36], although it occurred there at $Wi \approx 13$, whereas it occurs here for only $Wi \approx 0.395$, when the rest of the material properties are those of the 0.1% Carbopol solution of Lopez et al. (Table 2). Clearly, plasticity has contributed to the instability introducing it at a much lower value of fluid elasticity. Furthermore, yielded and unyielded areas are completely unsettled, changing with time in location, size and shape. The travelling waves inside the narrow channel become irregular. All these are indications of a strongly varying velocity and stress fields reminding us of turbulent conditions due to elasticity under creeping flow.

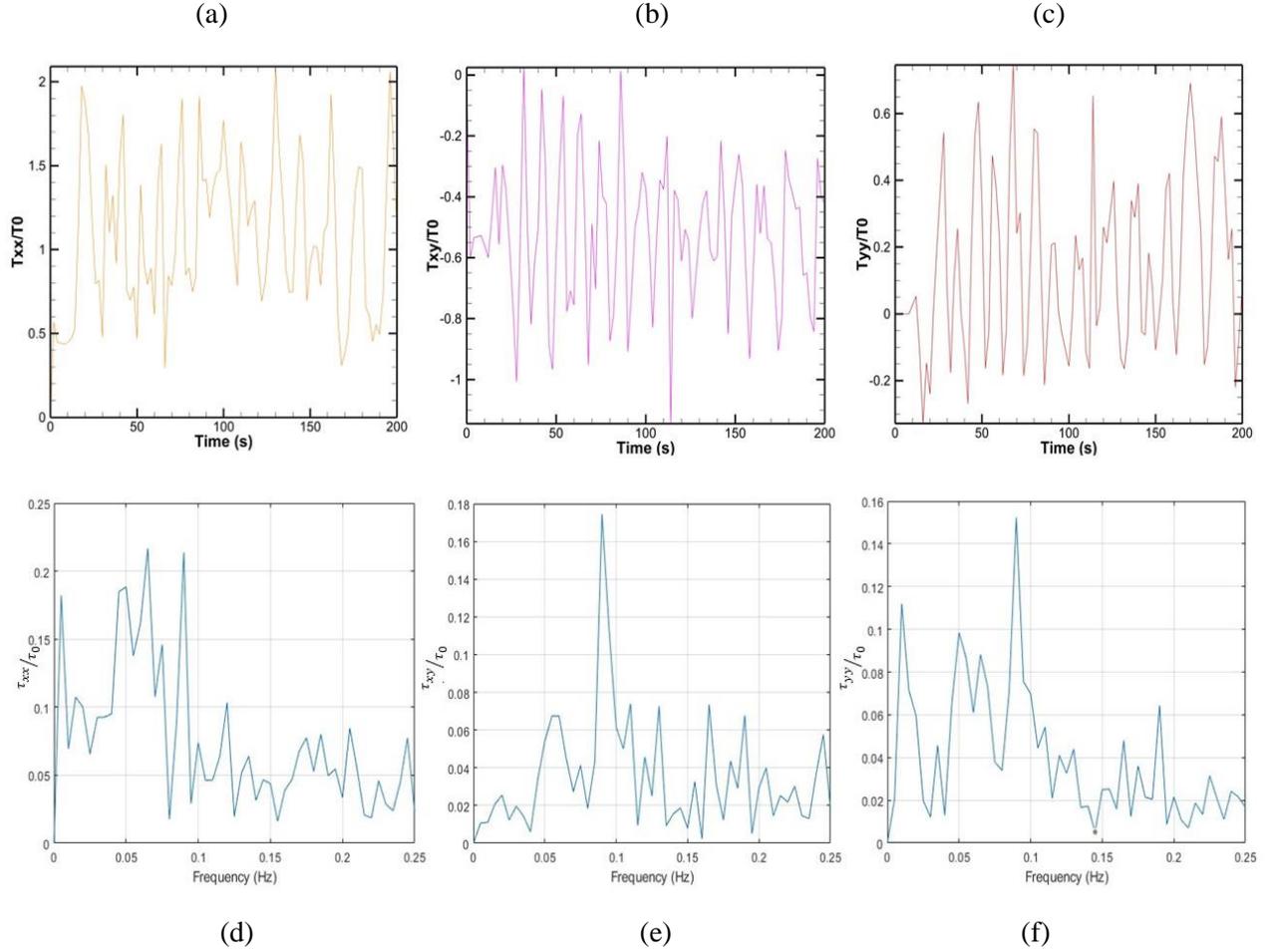

Figure 18. Evolution of nondimensional (a) $\frac{\tau_{xx}}{\tau_0}$ (b) $\frac{\tau_{xy}}{\tau_0}$ (c) $\frac{\tau_{yy}}{\tau_0}$ during the whole solution time for $G = 3\ Pa$ at the probe point $(\frac{x}{w}, \frac{y}{w}) = (5, 0.5)$. The FFT frequency analysis for (d) $\frac{\tau_{xx}}{\tau_0}$ (e) $\frac{\tau_{xy}}{\tau_0}$ (f) $\frac{\tau_{yy}}{\tau_0}$.

Figures 18 (a-c) illustrate the evolution of the three stress components ($\frac{\tau_{xx}}{\tau_0}, \frac{\tau_{xy}}{\tau_0}, \frac{\tau_{yy}}{\tau_0}$) from the beginning of the simulation until the final time ($t = 200s$). Both $\tau_{xx}$ and $\tau_{yy}$ exhibit oscillations with increased amplitude and multiple frequencies, displaying less smooth behavior compared to the $G = 4\ Pa$ case. Similar chaotic patterns are evident in the shear stress plots. Moreover, $\tau_{xx}$ remains always positive and has two dominant frequencies 0.065 Hz and 0.09, which are smaller than the frequency observed in the $G = 4\ Pa$ case. The dominant frequencies for $\tau_{yy}$ and $\tau_{xy}$ are 0.09 Hz, with the former assuming mostly positive values and the latter exclusively negative ones. The velocity evolution and corresponding frequencies are presented in Appendix D.

At the same time instances we present in Fig. 19 the velocity fields of the most disturbed flow of $G = 3\ Pa$. In the same panels, we depict the path lines with black continuous lines and arrows to denote the flow direction. To calculate them, we initially select a random set of positions in the entrance plane and near the salient corners. Then, we seed a single, massless particle from each point and monitor its trajectory with the velocity field at the same time instant. Due to this procedure to generate them and the transient nature of the flow, we do not produce either closed loops or intersecting lines near the salient corners. The corresponding stress components and pressure are given in Appendix B.

$t = 50\ s$

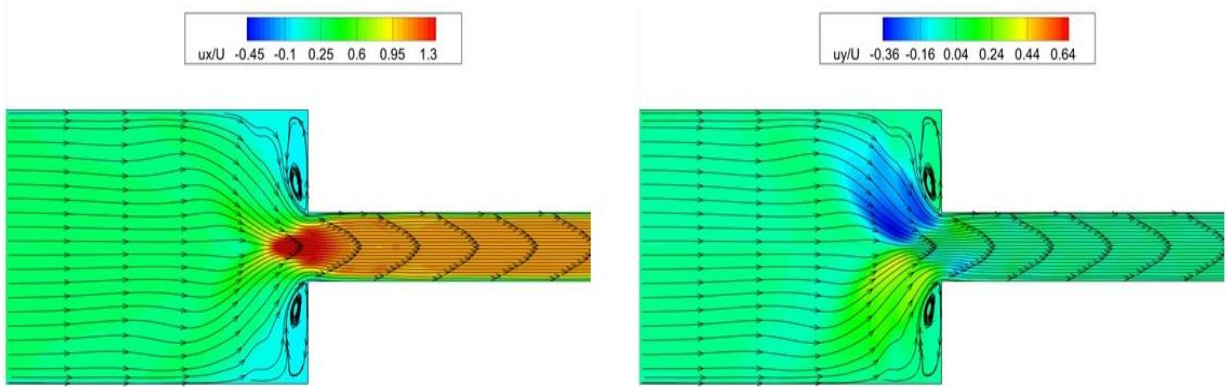

$t = 100\ s$

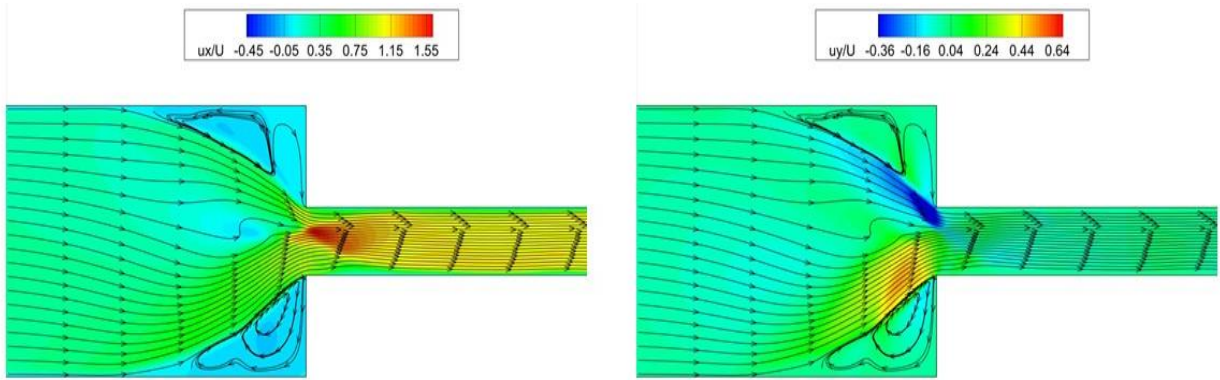

$t = 150\ s$

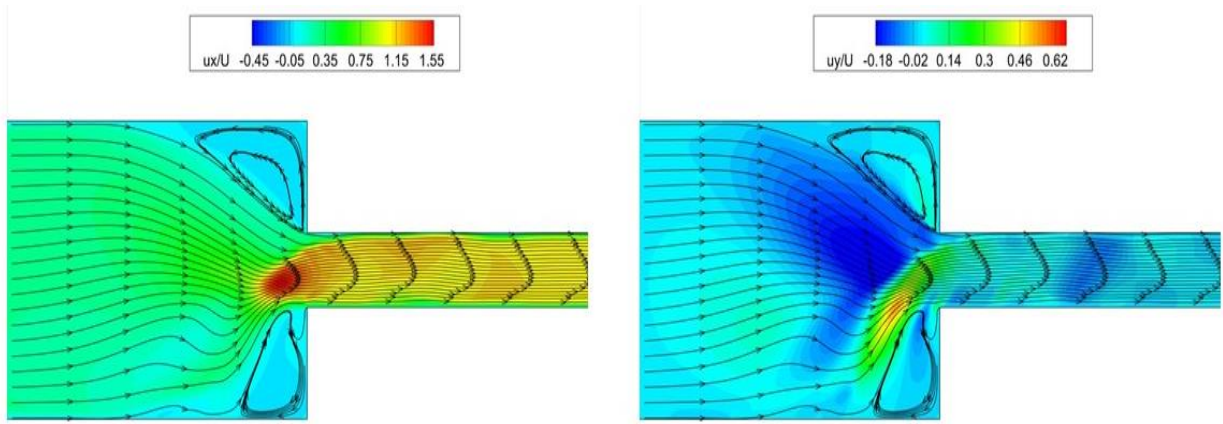

$t = 200\ s$

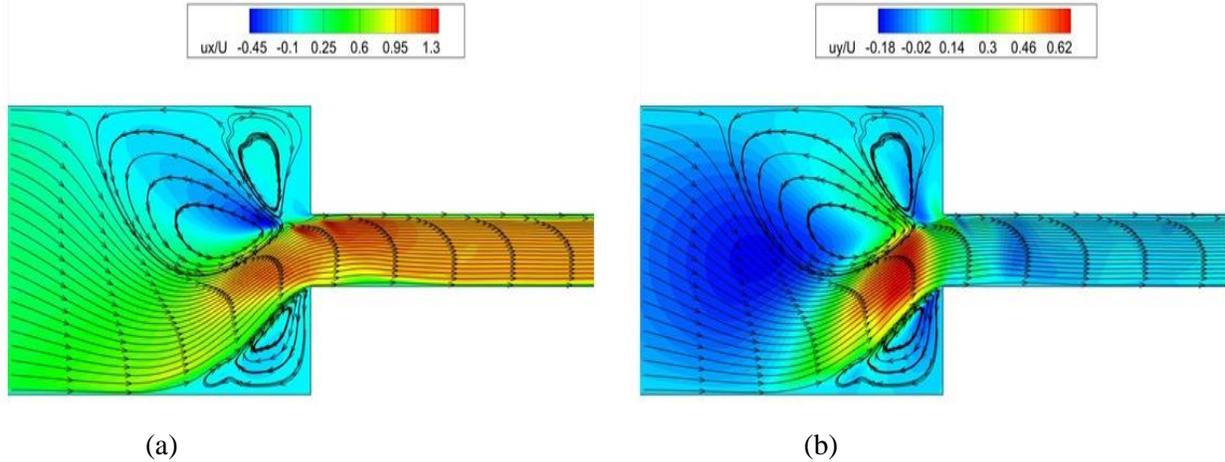

(a) (b)

Figure 19. Evolution of $(a) \frac{u_x}{U}$, $and$ $(b) \frac{u_y}{U}$, for $G = 3\ Pa$, while the rest of the parameters are those of the 0.1% Carbopol solution of Lopez et al. (Table 2)

At the first snapshot, $t = 50\ s$, both velocity components indicate that there is a plane of symmetry in the flow, in agreement with Fig. 17. The $u_x$ velocity is highest at the entrance of the contraction, while the magnitude of the $u_y$ velocity is highest at an angle above and below the entrance. The flow in the salient corner is rather weak and, because of its transient nature, the path lines do not depict a permanent recirculation pattern, but a continuously evolving path. This path is flattened against the wall by the stronger incoming flow, instead of having the previous triangular-like shape. The path lines just out of this pattern and in the bulk of the flow before the contraction are somewhat wavy. After the entrance to the narrow channel, the path lines become straight.

Subsequently, at $t = 100\ s$, the isolated flow in the salient corner extends to a much larger area upstream and becomes asymmetric with respect to the midplane, although the flow there remains weak. The path lines in the upper half seem to form two nearly closed counter-rotating loops, the first one somewhat away from the opposing wall and the second one nearly parallel to it. The path lines in the lower half form a triangular loop closer to the salient corner and a smaller one away from it. The maximum of the $u_x$ velocity is higher than previously, it is located inside the contraction and is clearly asymmetric as well, because it is closer to the upper wall. The path lines just ahead of the entrance to the contraction are even more corrugated than before.

Next, at $t = 150\ s$, a single loop appears again in each corner, and both are closer to the opposing wall, but remain asymmetric, as the upper one covers a larger area. In the main flow, the path lines are more wavy in the lower part of the wide channel and all along the narrow channel. The maximum of the $u_x$ velocity has moved back out of the narrow channel and now it is closer to the lower wall. Finally, at $t = 200\ s$, two distinct recirculations appear near the upper wall covering a very large area ahead of the salient corner. The recirculation in the lower salient corner covers a much smaller area. Their very different areas force the path lines in the central part of the flow to bend downward substantially and enter the contraction at an angle. The $u_x$ velocity is large both before and after the contraction entrance but is maximized inside it.

The flow asymmetry predicted for $G = 3\ Pa$ reminds us of the flow asymmetry reported for viscoelastic fluids when the $Wi$ increases in the same planar contraction [36, 47]. In principle, a purely elastic (i.e., inertialess) flow instability resulting from the coupling of the curved path lines near the contraction and the normal elastic stresses has led to this unsteady flow, according to the well-known criterion proposed by McKinley and co-workers [39, 40]. The need to compute even steady flows of EVP materials via transient simulations and the permanently transient nature of the flow do not allow determining streamlines and the stream function, and hence, the inception of this instability using either this criterion or performing a rigorous stability analysis. Overall, there is no sign that the transient evolution and the unsteadiness of the flow will cease, on the contrary, it seems to intensify with time.

## 4.4 Oscillatory or transient evolution when $\tau_o$ increases above a critical value

Next, we examine the effect of increasing the yield stress. We begin with $\tau_o = 18\ Pa$ in Fig. 20. We observe that even at $t = 200\ s$ the flow retains its plane of symmetry but remains transient. This is more clearly seen in the corresponding video in SM (https://veed.io/view/5c038e7f-fff2-4d76-bd94-3e1ac3ebe0b3). A thin unyielded finger is formed near the end of the thin yielded region next to the wall of the wide channel. The finger elongates, disappears, and reappears with time, but the shape of the front of the unyielded area in the wide channel seems to have stabilized. The yield surface in the narrow channel is smooth and its front in the entrance of the contraction is rather flat. Considering that this is the first transient solution with increasing $\tau_o$, it should resemble the first transient solution with decreasing, $G$, i.e. $G = 4\ Pa$, but except for the gross characteristics of the yield surfaces, it does not have any resemblance with the several details mentioned above.

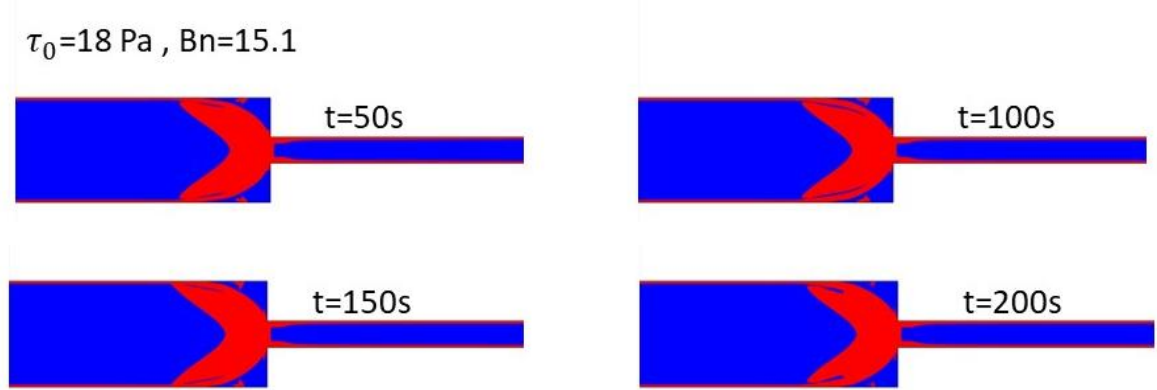

Figure 20. Evolution of the yielded (red) and unyielded (blue) zones for $\tau_o = 18\ Pa, (Bn = 15.1)$, while the rest of the parameters are those of the 0.1% Carbopol solution of Lopez et al. (Table 2). The yield strain of this case is $\varepsilon_y = 0.445$.

Increasing the yield stress to $\tau_o = 20\ Pa$ in Fig. 21, we observe that the width of the unyielded fingers has increased, and as they tend to increase in length, they split in more fingers downstream in each snapshot. The front of the unyielded area in the wide channel keeps varying turning from blunt and convex to pointed and concave and back (see the corresponding video https://veed.io/view/fc246bdf-c3f9-4b0c-94b5-0205dd2316fa). The flow remains symmetric. The yield surface in the narrow channel remains smooth, but its front in the contraction entrance keeps changing from flatter to pointed. No traveling waves can be detected as they did with decreasing values of $G$.

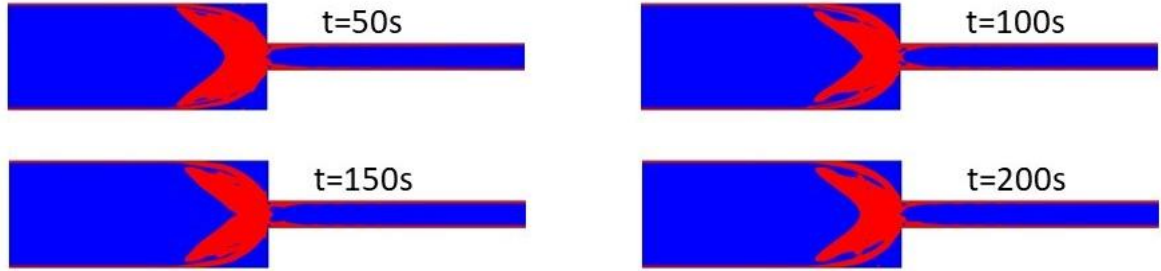

Figure 21. Evolution of the yielded (red) and unyielded (blue) zones for $\tau_o = 20\ Pa, (Bn = 16.78)$, while the rest of the parameters are those of the 0.1% Carbopol solution of Lopez et al. (Table 2). The yield strain of this case is $\varepsilon_y = 0.495$.

The case with $\tau_o = 21\ Pa$ is different from the previous one only in that the fingers increase in size and breakup into more unyielded areas, which lose their plane of symmetry at $t = 200\ s$, so we do not present it. The same trend continuous with $\tau_o = 22\ Pa$, except that the plane of symmetry is lost after $t = 100\ s$ (https://veed.io/view/b0c2cba9-a090-4efa-84dc-8e7d665b067f). So, we opted to report the case with the highest yield stress we examined, $\tau_o = 25\ Pa$, in Fig. 22, because it includes the most corrugated yield surfaces even from the first snapshot. We observe that several smaller unyielded areas appear inside the yielded area ahead of the contraction entrance. The flow and the yielded surfaces become nonsymmetric from the very beginning (https://veed.io/view/47503892-4815-4e1b-9a55-8359bf1eaa49). This asymmetry keeps growing and affects even the yield surface inside the narrow channel.

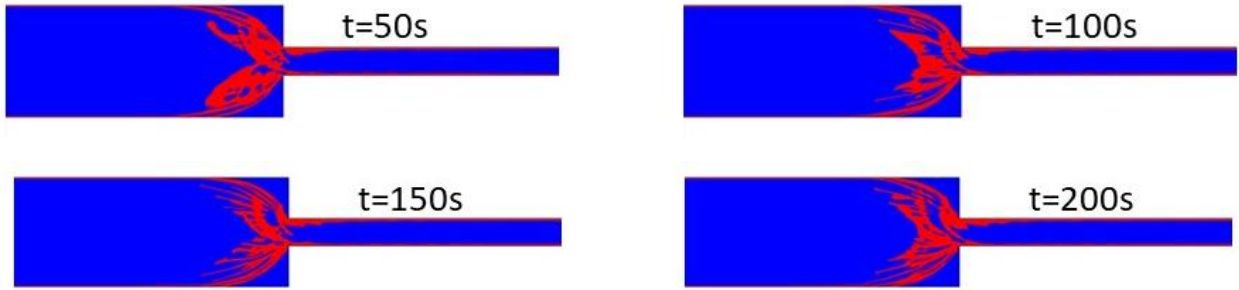

Figure 22. Evolution of the yielded (red) and unyielded (blue) zones for $\tau_o = 25\ Pa$, while the rest of the parameters are those of the 0.1% Carbopol solution of Lopez et al. (Table 2). The yield strain of this case is $\varepsilon_y = 0.619$

Despite the very convoluted yield/unyielded areas, the corresponding velocity field is quite smooth as seen in Fig. 23, whereas the stress and pressure fields are quite irregular (see Appendix C) leading to this form of the yield surfaces.

$$t = 50\ s$$

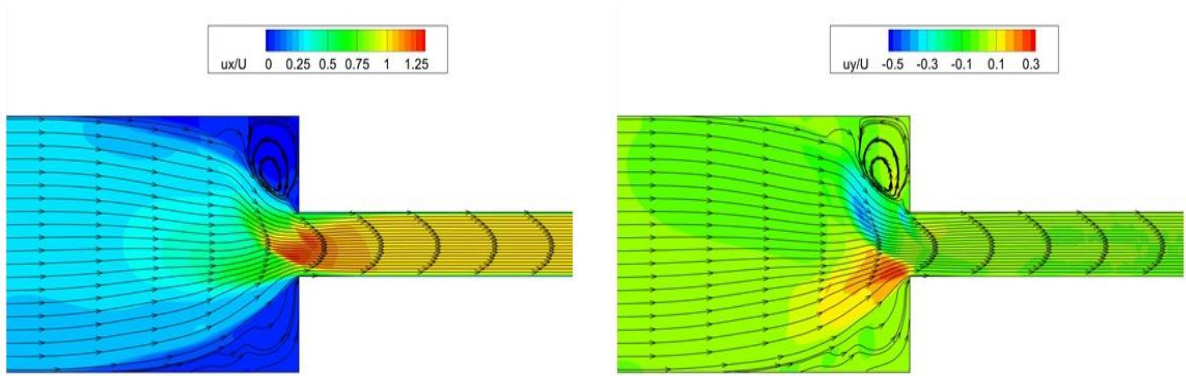

$t = 100\ s$

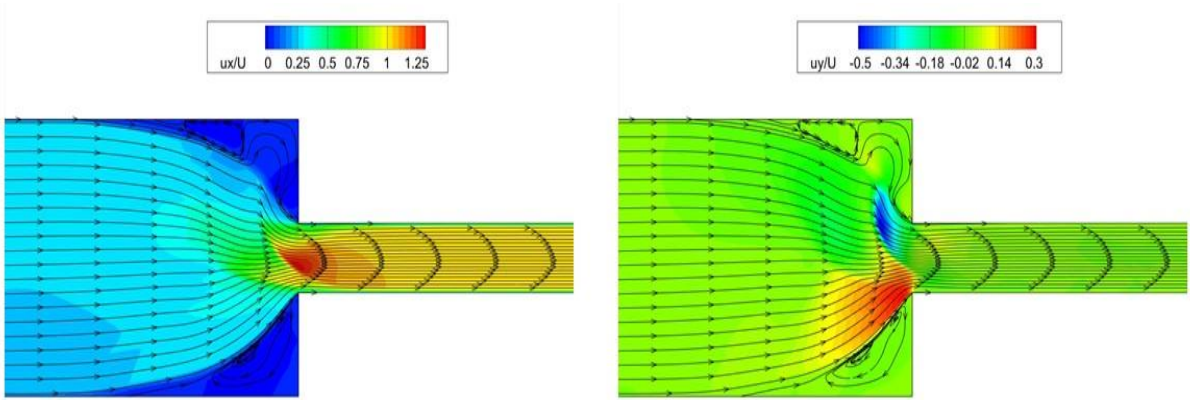

$t = 150\ s$

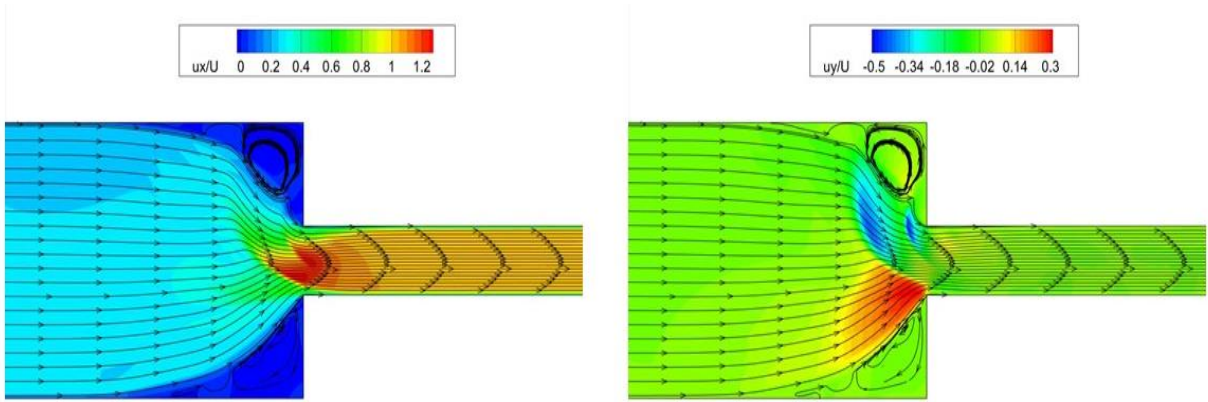

$t = 200\ s$

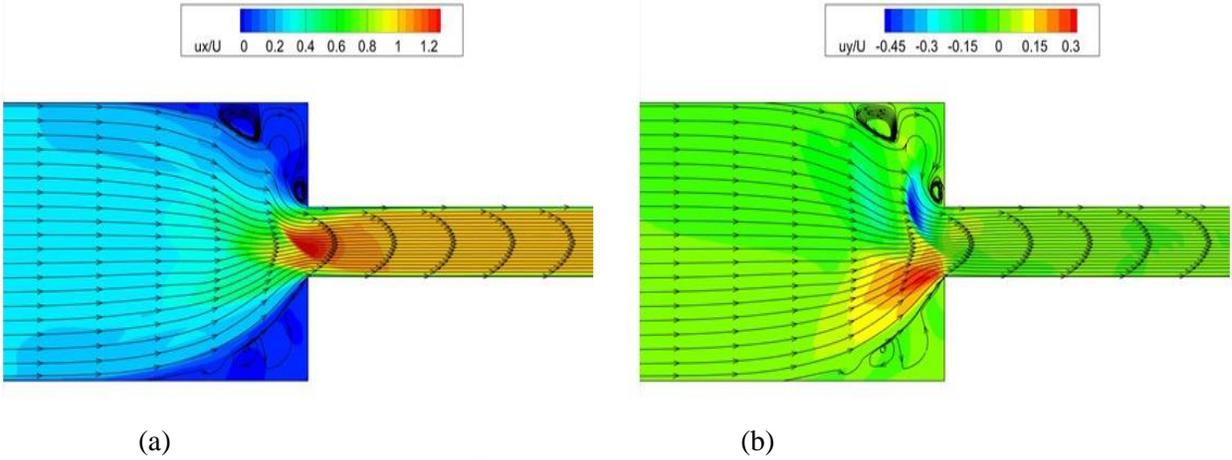

(a) (b)

Figure 23. Evolution of $(a)\frac{u_x}{U}$, and $(b)\frac{u_y}{U}$, for $\tau_o = 25\ Pa$, while the rest of the parameters are those of the 0.1% Carbopol solution of Lopez et al. (Table 2)

One may wonder why increasing the yield stress intensifies the elastic instability instead of decreasing it, via the increase of the effective viscosity. The reason for this is that increasing $\tau_o$ decreases monotonically the thickness of the yielded film next to the wall, primarily in the wide channel, $Y_1/w$, as can be seen in Table 5 and was observed in Fig. 12b. Decreasing the elastic modulus has the same effect as can be seen in the same Table and Fig. 13b, but the film thickness is about twice as large in the range of $G$ than it is in the range of $\tau_o$, where instability is initiated and examined herewith. As we mentioned already, the former is a well-known result of increasing $\tau_o$, whereas the latter is caused by the delayed material yielding caused by its higher elastic deformation prior to yielding.

| $\tau_o\ (Pa)$ | 15 | 18 | 20 | 21 | 22 | 25 |
|---|---|---|---|---|---|---|
| $Y_1/w$ | 0.26 | 0.185 | 0.15 | 0.136 | 0.13 | 0.089 |
| $Y_3/w$ | 0.22 | 0.20 | 0.182 | 0.193 | 0.18 | 0.1852 |
| $X/w$ | 3.86 | 3.614-3.66 | 2.77-3.7 | 2.239-3.513 | 2.5-3.65 | 2.21-3.55 |
| $\varepsilon_y = \tau_o/(G = 40.42\ Pa)$ | 0.371 | 0.445 | 0.495 | 0.520 | 0.544 | 0.619 |
| $G\ (Pa)$ | 3 | 4 | 5 | 10 | 20 | 40.42 |
| $Y_1/w$ | 0.28 | 0.316 | 0.35 | 0.57 | 0.75 | 0.8 |
| $Y_3/w$ | 0.25-0.5 | 0.32-0.6 | 0.46 | 0.42 | 0.42 | 0.41 |
| $X/w$ | 1.9-3.1 | 2.18-3.4 | 3.16-3.33 | 4.07 | 4.58 | 5.03 |
| $\varepsilon_y = (\tau_o = 4.71\ Pa)/G$ | 1.57 | 1.178 | 0.942 | 0.471 | 0.236 | 0.117 |

Table 5. Effect of the yield stress and the elastic modulus on the thickness of the film in either the wide $Y_1/w$ or the narrow channel $Y_3/w$, on the distance of the front of the unyielded material from the contraction entrance $X/w$, and, on the yield strain $\varepsilon_y = \tau_o/G$.

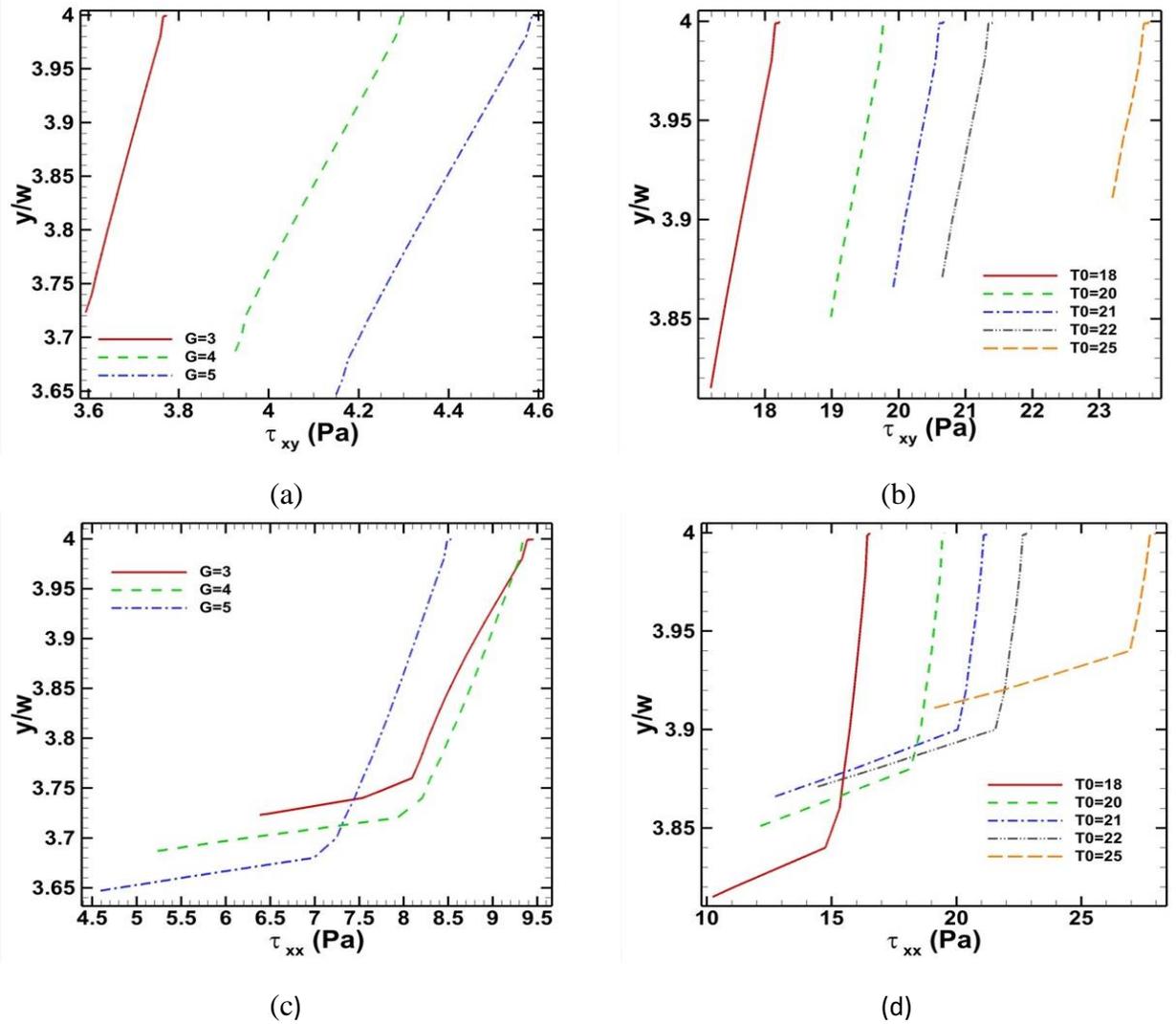

Figure 24. The shear stress (a) and (b) and normal stress (c) and (d) at the yielded film next to the wall ($\frac{x}{w} = -10$) of the wide channel as a function of the distance from the wall, which is located at ($\frac{y}{w} = 4$) up to the end of the yielded film, which depends on the parameter value. In (a) and (c) $G = 3, 4, 5\ Pa$ with $\tau_o = 4.71\ Pa$; in (b) and (d) $\tau_o = 18, 20, 21, 22, 25\ Pa$, with $G = 40.42\ Pa$, while the rest of the parameters are those of the 0.1% Carbopol solution of Lopez et al. (Table 2).

This common decrease in film thickness by increasing $\tau_o$ or decreasing $G$ leads to an increase in the magnitude of the stress in this film, but with different contributions from the stress components. This can be easily seen in Fig. 24. When instability is generated by increasing $\tau_o$, its value (indicated in the inset in Fig. 24(b) and 24(d)) is larger than when it is generated by decreasing $G$ ($\tau_o = 4.71\ Pa$). Then, $\tau_{xy}$ is higher than $4.71\ Pa$ and contributes about the same with $\tau_{xx}$ to the von Mises criterion for yielding and both increase as the film thickness decreases. Conversely, when instability is generated by decreasing $G$, its value (it assumes the values indicated in the inset in Fig. 24(a) and 24(c)), is smaller than $G$ in Fig. 24 (b), which is as high as $40.42\ Pa$. Then, $\tau_{xy}$ contributes much less than $\tau_{xx}$ to the von Mises criterion for yielding and only the latter increases as the film thickness decreases. It is noteworthy that in all cases $\tau_{xy}$ increases almost linearly within the thin film as the wall is approached, whereas $\tau_{xx}$ also increases linearly closer to the wall, but sharply turns to smaller values as the unyielded area is approached.

Asymmetric flow can manifest in both high $Bn$ and $Wi$ flows [48]. The effects of the yield stress and the material elasticity on this and other flows of EVP materials can be examined more concisely by introducing the yield strain parameter, $\varepsilon_y = \tau_o/G$, which depends only on material properties and increases when either $\tau_o$ increases or $G$ decreases. It is noteworthy that $G$ arises in the denominator of the elastic term of the constitutive relation, whereas $\tau_0$ in the numerator of the viscoplastic term. As far as we know, the yield strain was first introduced in Varchanis et al. [23] and then often used to describe phenomena in EVP fluids [49, 50]. An increase in yield strain makes the unyielded material more elastic, as manifested by the increase in the normal elastic stress components. These significantly contribute to the von Mises criterion, causing material yielding at lower shear stress values than usual. This is depicted in Fig. 24, where the thin film yields at lower shear stress values when $G$ increases.

When the solvent viscosity is negligible, (as in this study with $\eta_s = 0.01\ Pa\ s$, but is included only to retain the ellipticity of the momentum balance), the product of the Weissenberg number with the Bingham number results in yield strain, $\varepsilon_y = \frac{\tau_0}{G} \approx Wi * Bn$. Also, this product was identified as the main driving force for flow asymmetries instead of either one of $Bn$ or $Wi$ in other previous studies [45, 46]. On the other hand, the effect on the inception of the instability by decreasing $G$ is stronger than the effect by increasing $\tau_0$, because of the definition of the yield strain, $\varepsilon_y = \frac{\tau_0}{G}$. The base case is always the Lopez et al. material, which has $\tau_o = 4.71\ Pa, G = 40.42\ Pa$. Instability is observed when $\tau_0$ increases from 15 Pa to 18 Pa, this is an increase by $(3/40.42) * 100 = 7.4\%$ in $\varepsilon_y$, similarly instability is observed when $G$ decreases from 5 Pa to 4 Pa, this an increase by $(4.71/4 - 4.71/5) * 100 = 23.55\%$ in $\varepsilon_y$. In other words, decreasing $G$ by 1 Pa affects the denominator of $\varepsilon_y$, whereas increasing $\tau_0$ by 3 Pa affects the numerator of $\varepsilon_y$ while keeping everything else the same as in the base case. Consequently, even a slight decrease in the elastic modulus leads to a larger alteration in the stress components, than increasing the yield stress. All these may be observed in Table 5 as well. Closer examination of the values in this Table makes clear that the yield strain alone cannot determine transition to instability.

This increase in the magnitude of the stress when the film gets thinner detaches unyielded material from the main unyielded area forming unyielded fingers of increasing length and width, which may break inside the yielded area ahead of the contraction. This phenomenon coupled with increasing oscillation amplitude of the front of the unyielded material (indicated by the widening range of values assumed by $X/w$ which can be seen in Table 5) forces the path lines to become more distorted and adopt a smaller radius of curvature, which is known to induce elastic instability [39, 40]. The instability is more intense when the elastic modulus also decreases, as opposed to when the yield stress increases, according to the analysis in [39, 40]; compare Fig. 17 and 19 with Fig. 22 and 23.

## 5   Summary and Conclusions

In this study, the flow in the planar contraction (4:1) benchmark geometry is investigated with the Saramito-Herschel-Bulkley (SHB) elasto-viscoplastic model (EVP). The OpenFOAM software coupled with the RheoTool toolbox is used for the transient numerical simulations. It uses the Finite Volume Method to discretize the governing equations, in a mesh which is locally refined near singular points or where the solution is expected to vary more abruptly. First, convergence of the numerical method and our setup is determined by examining, with successively refined meshes, the flow of a viscoelastic fluid, which has been studied very extensively over many years. Solution at high Weissenberg numbers is achieved with the help of the log-conformation formulation. When the Oldroyd-B fluid model is used, two vortices arise close to re-entrant and concave corners, the so-called lip and corner vortices, respectively. Their size is a very sensitive measure of the accuracy of the solution, and we demonstrate excellent agreement with the results of prior studies.

Turning to EVP materials, we examined first the flow in this geometry of three well-characterized Carbopol gels of 0.09% and 0.1% concentrations. Despite the very similar concentrations, they exhibit quite different yield surfaces and flow fields. Five non-dimensional groups may affect the simulations in general: Reynolds, Weissenberg, Bingham, yield strain, and the solvent to solution viscosity ratio. The Reynolds number remains in all simulations much smaller than one. Then, a parametric study is performed by varying the four rheological parameters ($\tau_0, G, n, k$) of the SHB model one at a time, while keeping the rest of the values to those of one of the two Carbopol 0.1% solutions. The numerical results are presented mainly in terms of the yield surfaces. Their shapes are characterized by four lengths defined in Fig. 10. Both $n$ and $k$ affect four of the dimensionless groups. Their more significant effect is that the thickness of the yielded film next to the wall of the wide channel and the distance of the tip of the unyielded material there from the contraction entrance decrease, when $n$ increases or $k$ decreases.

By definition, $\tau_0$ affects the Bingham number proportionately, but no other dimensionless number. As expected, its increase expands the unyielded areas monotonically, as demonstrated also by all four lengths we defined to characterize them. Similarly, $G$ affects the Weissenberg number in an inversely proportional manner. Increasing $Wi$ also expands the unyielded areas, as demonstrated also by the lengths characterizing them, except for the film thickness inside the narrow channel, which remains nearly constant. This outcome stems from the fact that more elastic materials have higher resistance towards yielding than stiffer materials. Furthermore, a higher elasticity level relocates the unyielded areas in the flow direction.

All the results discussed above were obtained when a steady state was reached and when the Bingham and the Weissenberg numbers remained below a certain critical value. When their respective critical value was exceeded, the flow remained transient becoming either periodic with a decreasing amplitude and varying wavelength or chaotic. The mechanism leading to this elastic instability, when $Wi$ increased, is related to the coupling of increased fluid elasticity with instantaneous radius of curvature of the path lines, first reported in [39, 40]. However, a similar instability when the Bingham number is increased is quite unexpected, because under these circumstances the flow should tend to become smoother and hence steadier. A closer examination revealed that in fact increasing either one of these two dimensionless numbers decreases the thickness of the yielded film next to the wall of the wide channel, which in turn increases the magnitude of the stress locally and disrupts the nearby unyielded area by inducing unyielded fingers ahead of it. This disrupts further the path lines and causes this elastic instability. Having a similar mechanism, the inception of the instability by these two dimensionless numbers is synergistic, so the critical values we determined are not absolute, but depend on each other, this means for example, that the instability will arise at a lower $Wi$ if a material with a larger Bingham number is examined. Introducing the yield strain makes this description more concise.

It would be very interesting to examine this flow of EVP materials experimentally and determine the conditions leading to the reported instabilities. In the second part of this study, we will present the flow of EVP fluids around a confined cylinder.

**Appendix A:** The EVP mesh convergence study using the meshes detailed in Table 4

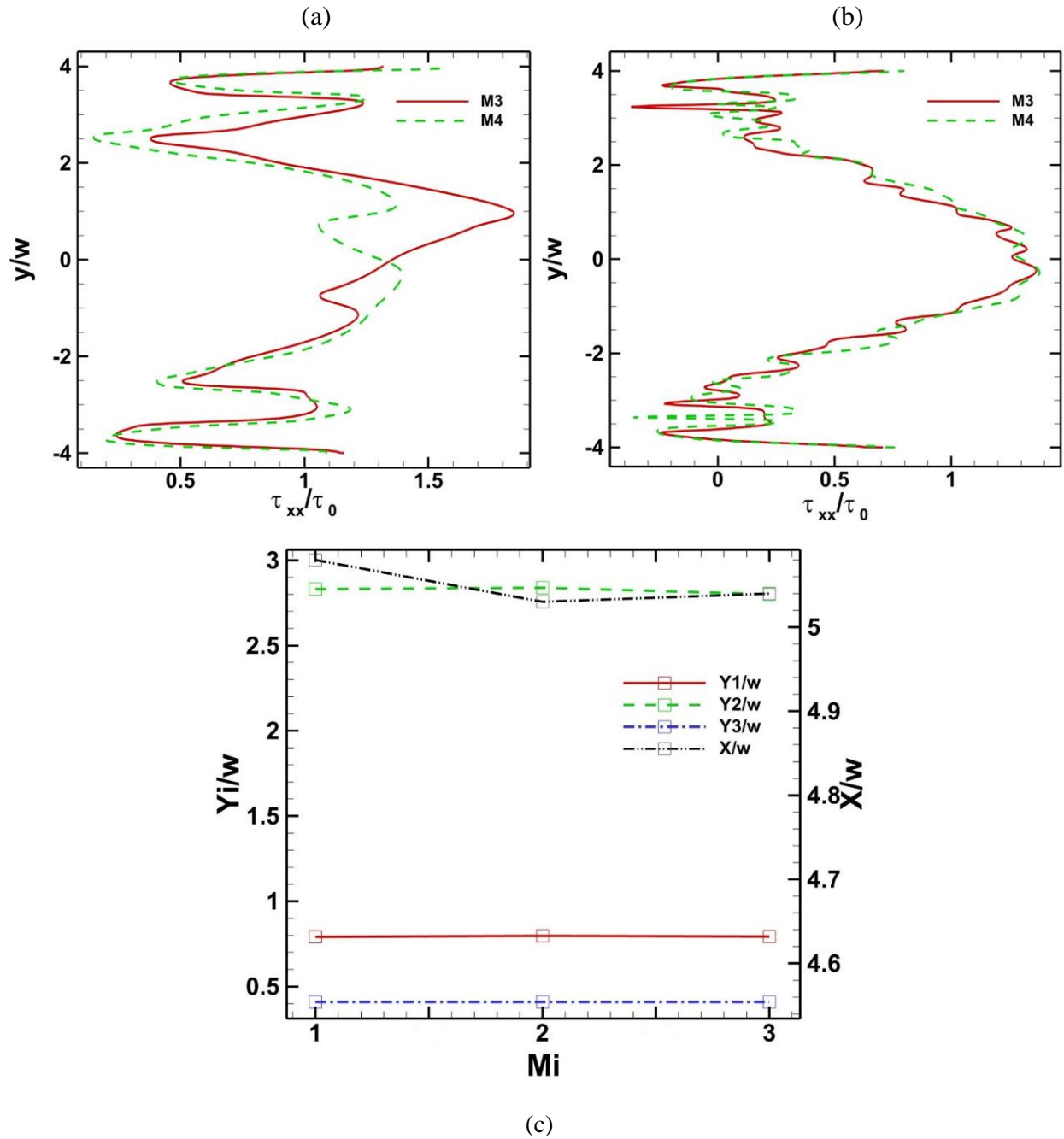

Figure 25. The dimensionless normal stress component ($\tau_{xx}/\tau_0$) along the $y$-axis at $\frac{x}{w} = -3$ for a) $G = 3\ Pa$ & b) $\tau_0 = 22\ Pa$. c) Convergence of the characteristic lengths for three meshes M1, M2, and M3 detailed in table 4 for the Lopez 0.1% Carbopol.

In Fig. 25, additional cases for mesh convergence are illustrated for conditions leading to either steady or transient solutions. Figs. 25a and 25b depict the variations in the normal stress component ($\tau_{xx}/\tau_0$) along the $y-axis$ at $\frac{x}{w} = -3$ for $\tau_0 = 22\ Pa\ \&\ G = 3\ Pa$ where transient behavior is observed. It is crucial to emphasize that in the proximity of the contraction plane at the centerline, the extensional stress component undergoes significant variations with slight changes in the mesh configuration. This observation aligns with what was noted in the mesh convergence study concerning the lip vortex size with Oldroyd-B fluids. However, Fig. 25b indicates a reasonable agreement for $\tau_0 = 22\ Pa$. All results are reported at the final time instant (t=200s). In Fig. 25c, the dependence on mesh refinement of the characteristic lengths for the yielded/unyielded region is presented, according to their definitions in Fig. 10a. Fig. 25c demonstrates a very good agreement for the three meshes used.

**Appendix B:** Evolution of stress fields and pressure when $G = 3\ Pa$

$t = 50\ s$

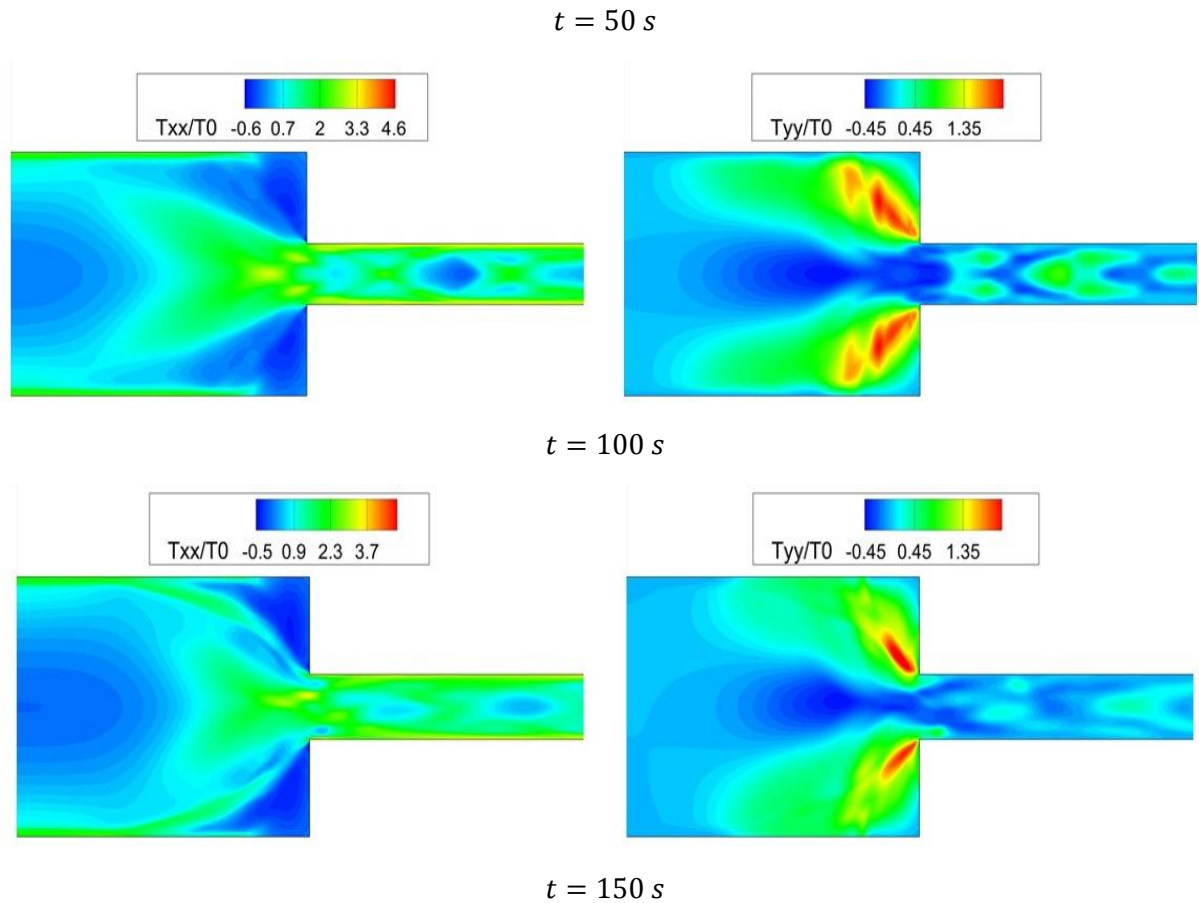

$t = 100\ s$

$t = 150\ s$

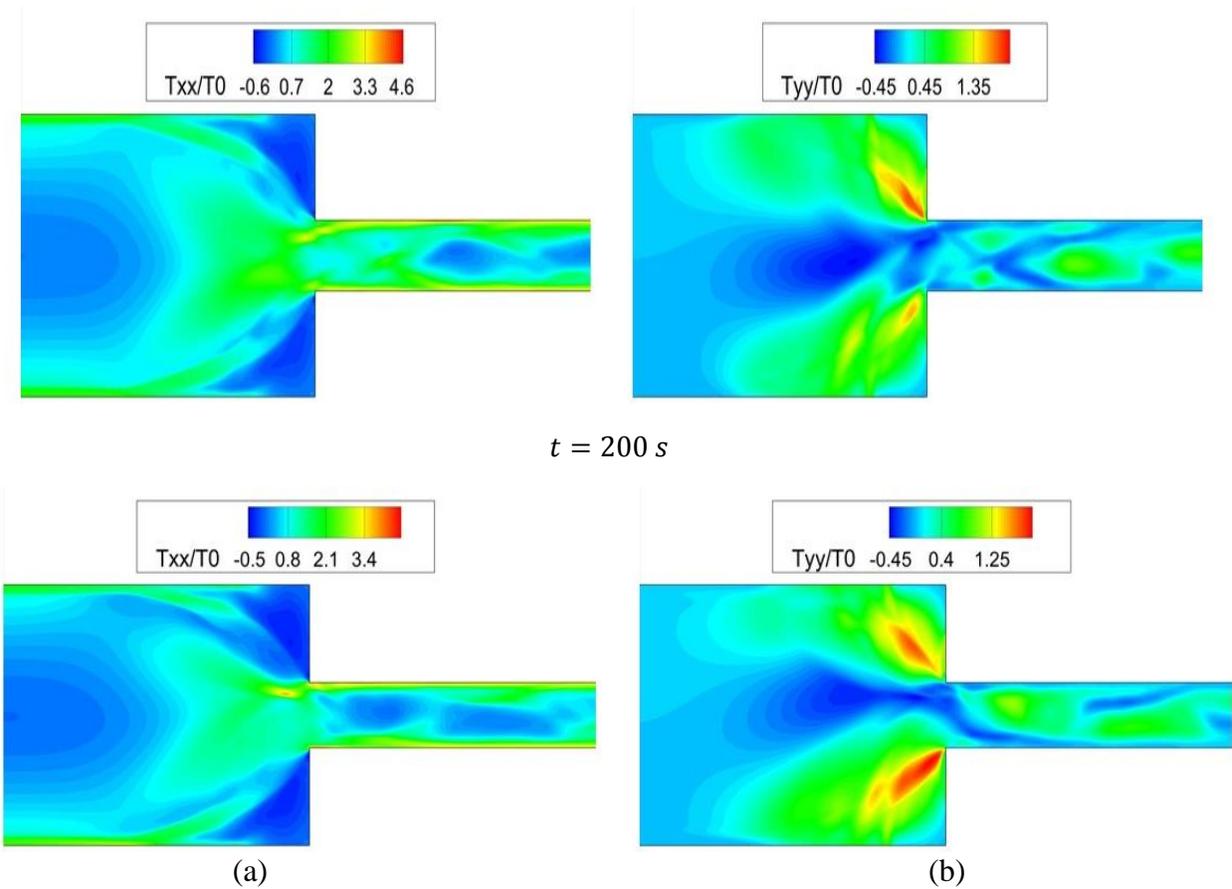

(a)                                                                       (b)

Figure 26. Evolution of $(a)\frac{\tau_{xx}}{\tau_0}$ $and$ $(b)\frac{\tau_{yy}}{\tau_0}$ for $G = 3\ Pa$, while the rest of the parameters are those of the 0.1% Carbopol solution of Lopez et al. (Table 2)

At the first snapshot in Fig. 26, both normal stress components are symmetric with respect to the midplane and display oscillations inside the narrow channel, with values that undergo significant fluctuations. These variations in normal stresses lead to rapid transitions between solidification and fluidization over very short distances. Subsequently, these fluctuations appear to decrease, become irregular and asymmetric, yet the abrupt transition between yielded and unyielded material is happening. They continue until the end of the simulations, becoming more irregular inside both channels. It may be noticed that they tend to have longer wavelengths, are more widely spaced, and the two normal stress components vary in opposite directions. In the proximity of both the upper and lower re-entrant corners, the intense but asymmetric $\tau_{yy}$ contributes to the ongoing fluctuations within the narrow channel.

$$t = 50\ s$$

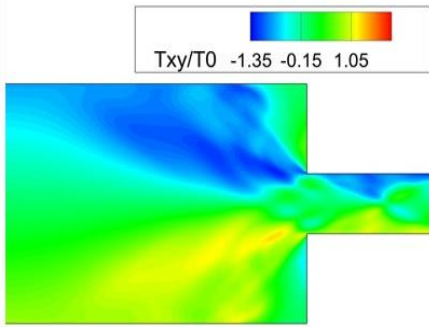 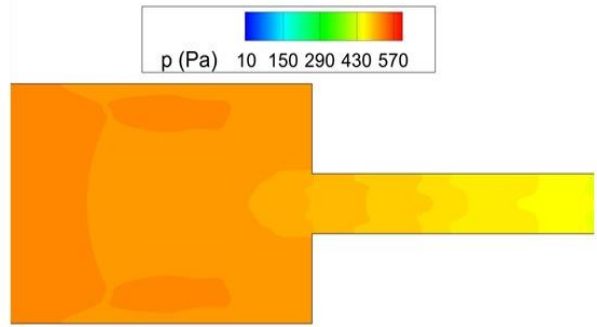

$t = 100\ s$

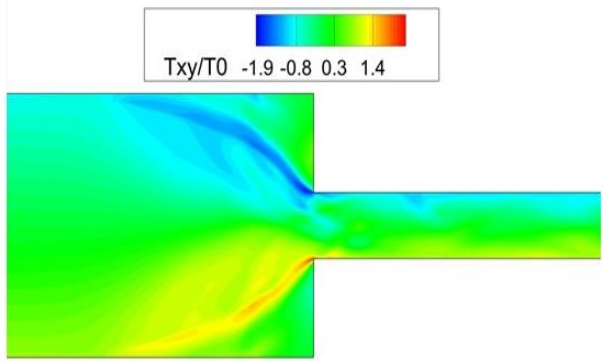 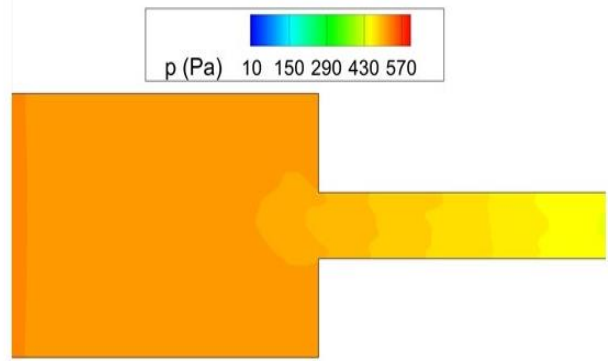

$t = 150\ s$

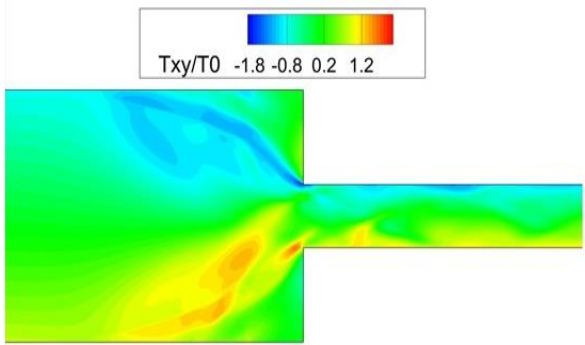 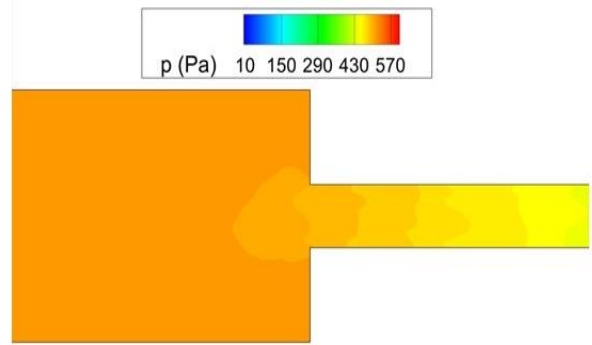

$t = 200\ s$

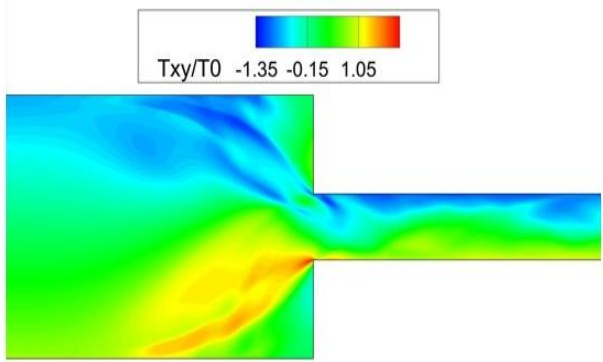 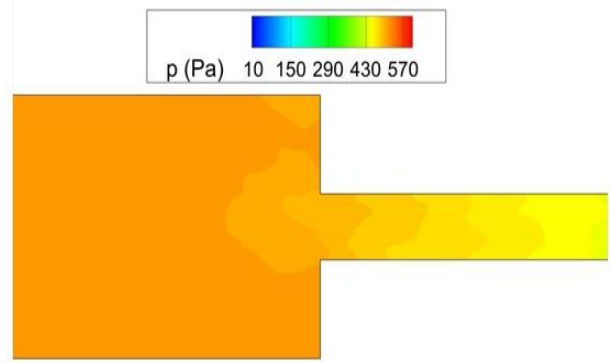

|  (a)  |  (b)  |

Figure 27. Evolution of $(a)\frac{\tau_{xy}}{\tau_0}$ and $(b)\ P$ for $G = 3\ Pa$, while the rest of the parameters are those of the 0.1% Carbopol solution of Lopez et al. (Table 2)

As expected, the shear stress exhibits larger values near the channel walls, displaying opposing slopes on either side (See Fig. 27). Furthermore, the pressure within the narrow channel registers notably lower and rather uniform values in comparison to the wider channel. At the first time instant, both have a plane of symmetry, like every other flow variable, which disappears subsequently. Then the shear stress fluctuates both near the walls and in the vicinity of the re-entrant corners. At $t = 150\ s$, the shear stress becomes more uniform, but at the final snapshot it becomes nonuniform again. The distribution of pressure aligns with the predictions from the earlier time instances.

## Appendix C: Evolution of stress fields and pressure when $\tau_o = 25\ Pa$

The most significant observation from Fig. 28 and 29 in comparison to the $G = 3\ Pa$ case, is that the intensity of fluctuations of all stress components and particularly of $\tau_{xx}$ and $\tau_{yy}$ has increased considerably.

$t = 50\ s$

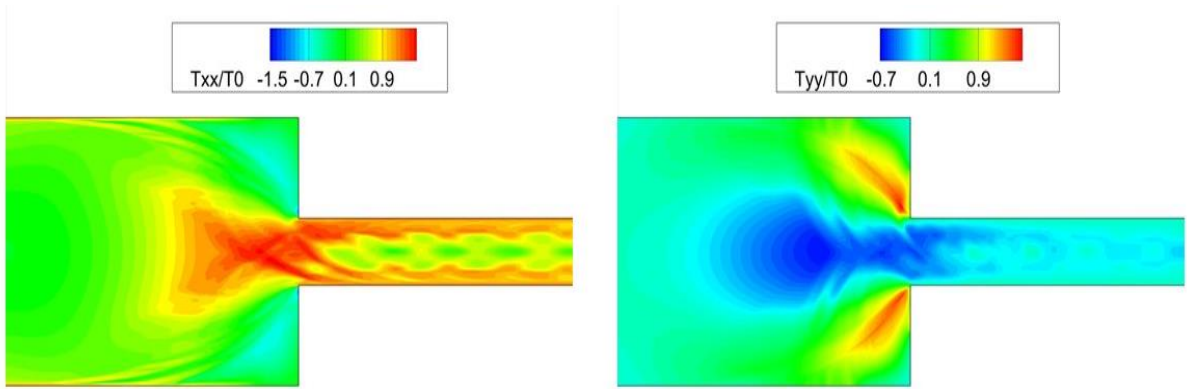

$t = 100\ s$

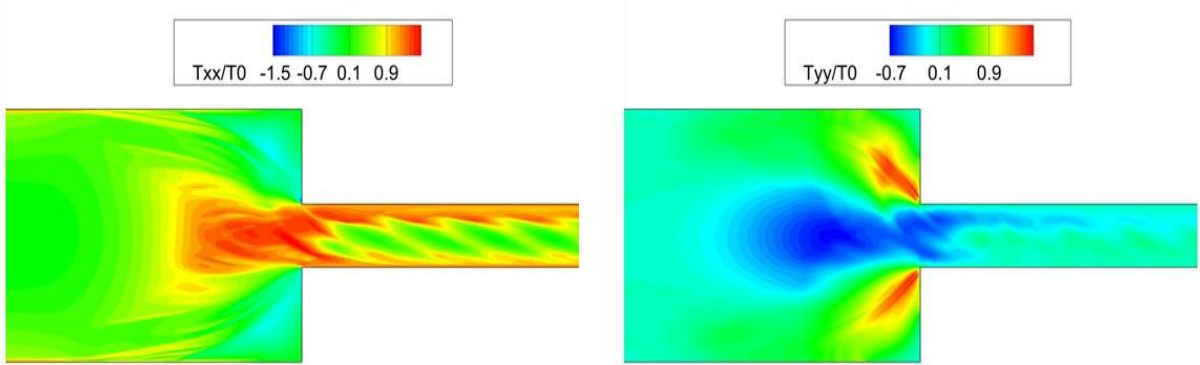

$t = 150\ s$

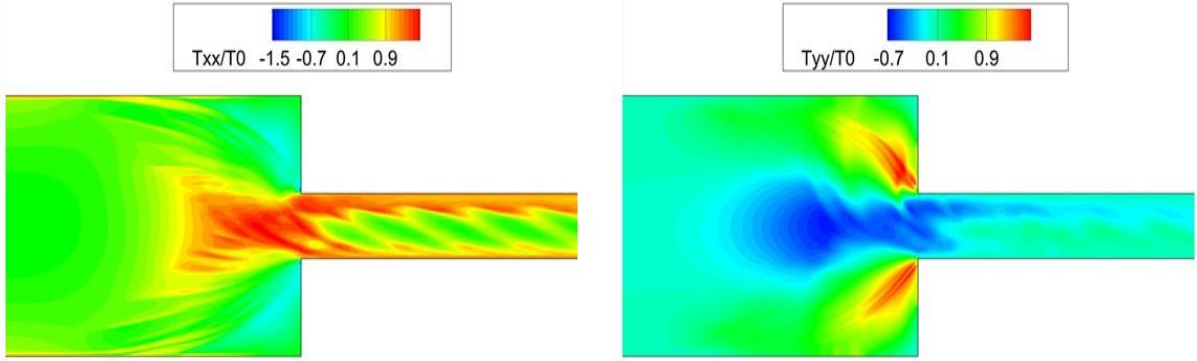

$t = 200\ s$

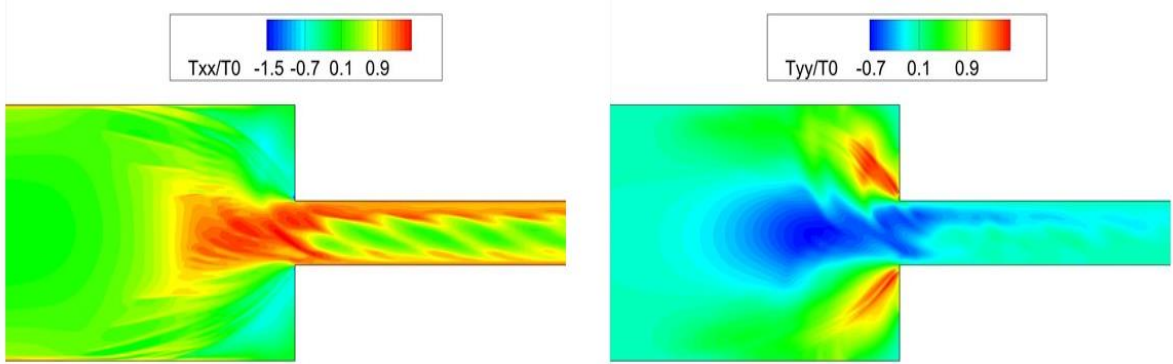

(a)                                      (b)

Figure 28. Evolution of $(a)\ \dfrac{\tau_{xx}}{\tau_0}\ and\ (b)\ \dfrac{\tau_{yy}}{\tau_0}$ for $\tau_o = 25\ Pa$, while the rest of the parameters are those of the 0.1% Carbopol solution of Lopez et al. (Table 2)

$t = 50\ s$

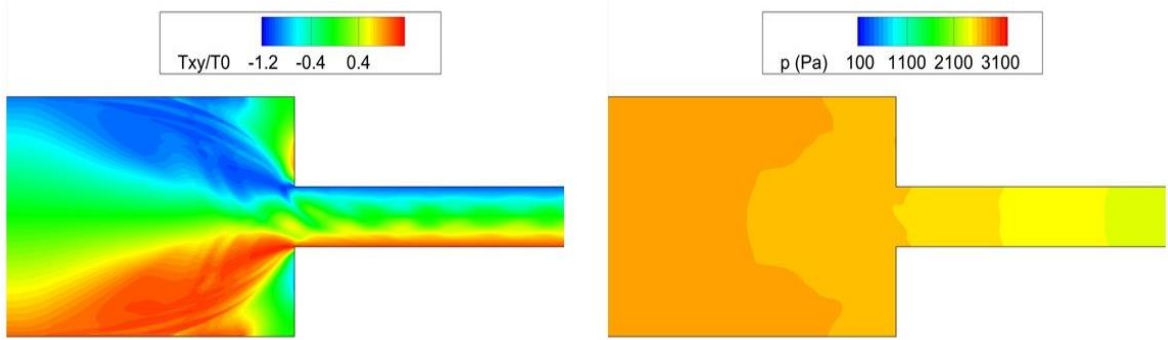

$t = 100\ s$

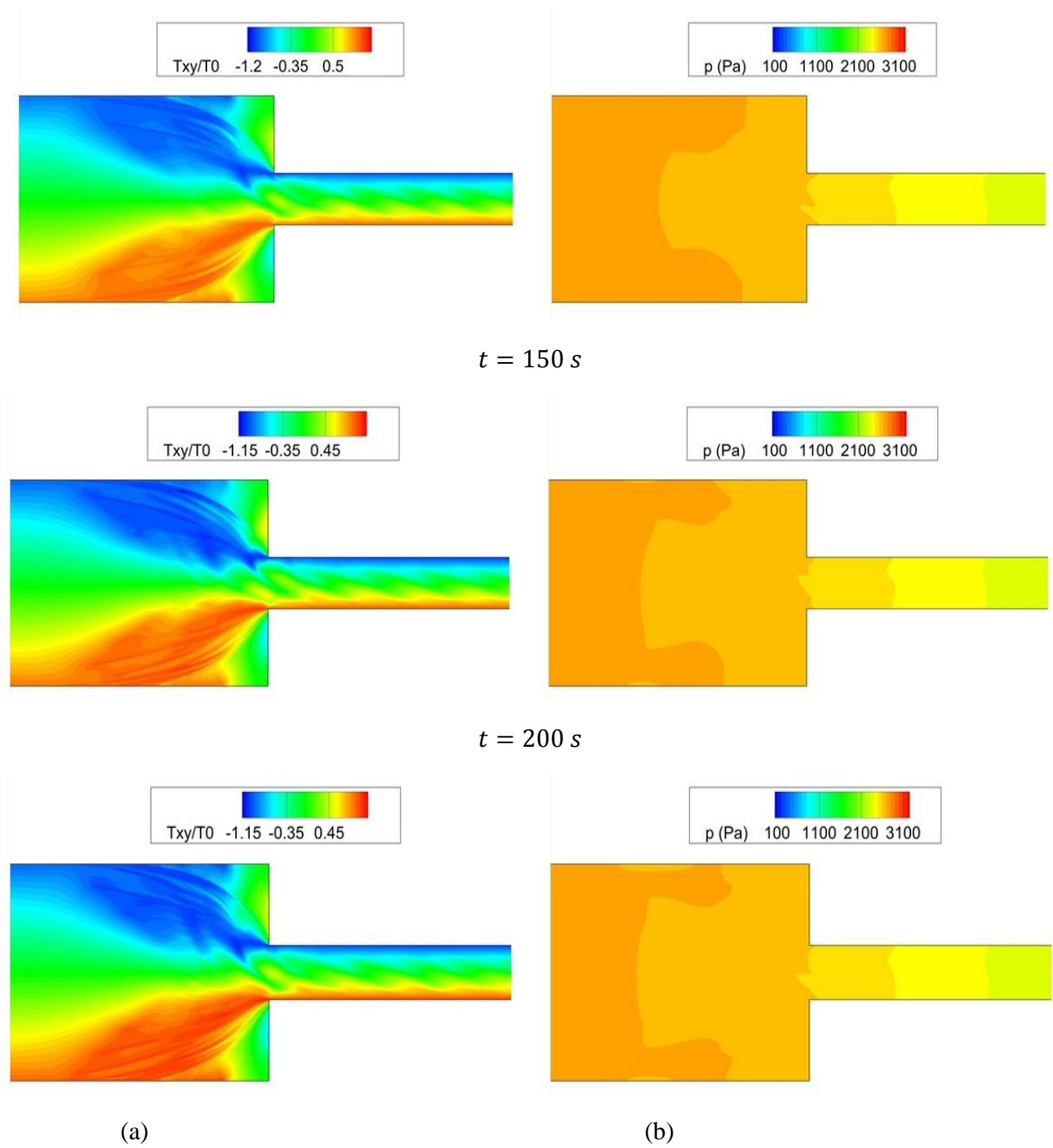

$t = 150\ s$

$t = 200\ s$

(a)          (b)

Figure 29. Evolution of $(a)\ \frac{\tau_{xy}}{\tau_0}$ and $(b)\ P$ for $\tau_o = 25\ Pa$, while the rest of the parameters are those of the 0.1% Carbopol solution of Lopez et al. (Table 2)

**Appendix D:** Evolution of velocity and the corresponding FFT analysis for $G = 3\ Pa$, $G = 4\ Pa$, and $\tau_o = 25\ Pa$.

(a) 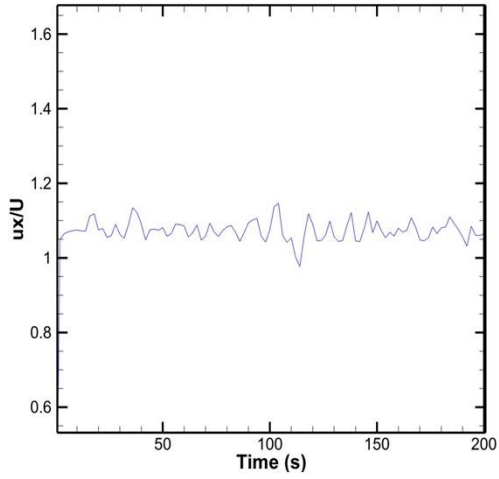 (b) 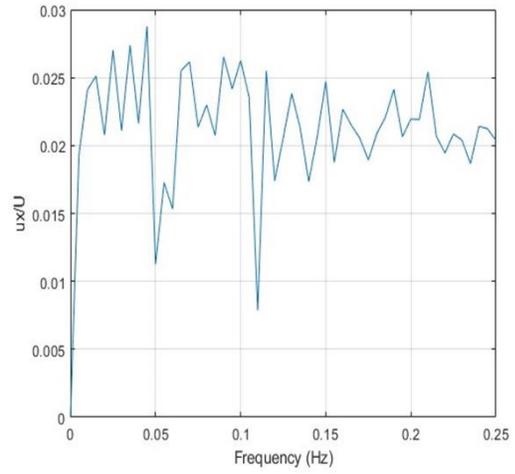

(c) 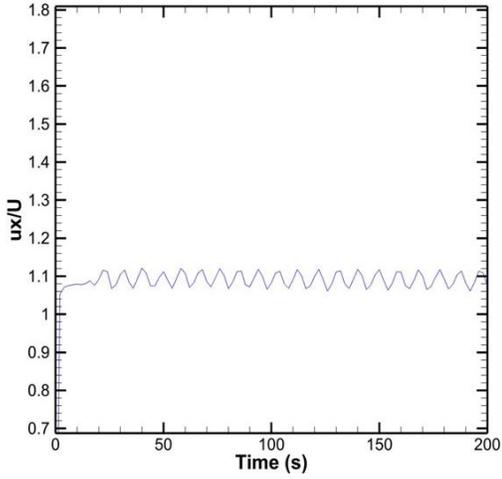 (d) 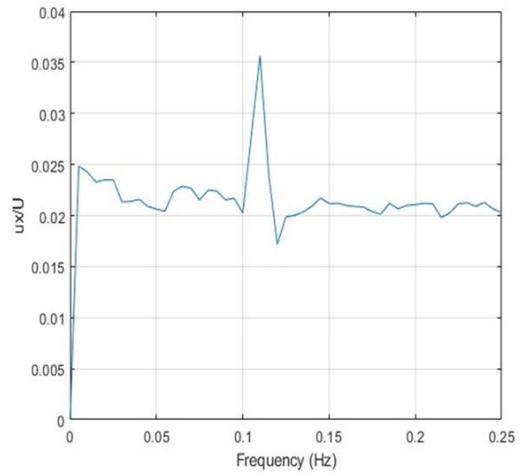

(e) 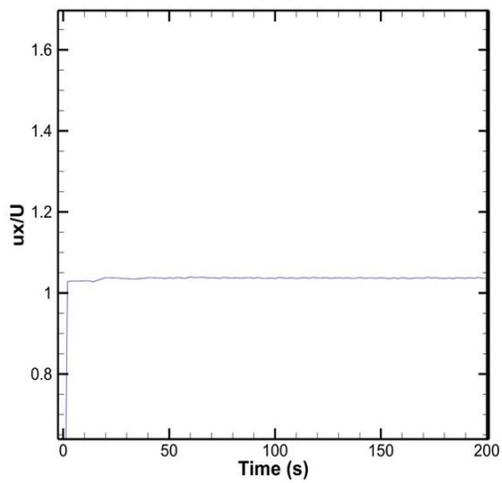 (f) 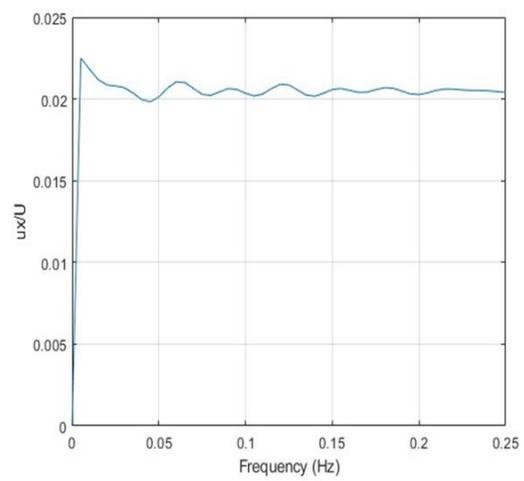

Figure 30. Evolution of $\frac{u_x}{U}$ for $(a)\ G = 3\ Pa\ (c)\ G = 4\ Pa\ (e)\ \tau_o = 25\ Pa$ during the whole solution time at the probe point $(\frac{x}{w}, \frac{y}{w}) = (5, 0.5)$. The FFT frequency analysis for $(b)\ G = 3\ Pa\ (d)\ G = 4\ Pa\ (f)\ \tau_o = 25\ Pa$.

Fig. 30 depicts the evolution of horizontal velocity in three transient cases. The variations in velocity observed during the simulation highlight that the most irregular patterns are observable for $G = 3\ Pa$. In contrast, $G = 4\ Pa$ exhibits regular patterns with a dominant frequency of 0.11. The third case involves $\tau_0 = 25\ Pa$, where the velocities display no variation in time, and no discernible frequency. This is contrast to the obvious variations of all stress components of the case in Fig. 29. These results demonstrate that the chaotic behavior of the velocity field is significantly more pronounced in flows with high $Wi$ compared to flows with high $Bn$. It is noteworthy to mention that the vertical velocity ($u_y$) values are approximately two orders of magnitude smaller compared to the horizontal ones ($u_x$).

## Acknowledgement

This research has received funding from the European Union´s Horizon 2020 research and innovation programme under the Marie Skłodowska-Curie grant agreement No 955605. The assistance by P. Moschopoulos in preparing certain plots is greatly acknowledged.

## Declaration of interests.

The authors report no conflict of interest.